\documentclass[journal]{IEEEtran}
\usepackage{amsmath,graphicx}
\usepackage{epstopdf}
\usepackage{latexsym}
\usepackage{amssymb}
\usepackage{cite}
\usepackage{ushort}
\usepackage{mathtools}
\usepackage{caption}
\usepackage{url}
\usepackage{pgfplots}
\usepackage{tikzscale}
\usepackage{booktabs}
\usepackage{ushort}
\usepackage{calrsfs}
\usepackage{makecell}
\usepackage{array}
\usepackage{textcomp}
\usepackage{calc}
\usepackage{tikz}
\usetikzlibrary{shapes,arrows}
\usetikzlibrary{calc}
\usetikzlibrary{positioning}
\usetikzlibrary{arrows}
\usetikzlibrary{arrows.meta}
\usetikzlibrary{arrows,scopes}
\usetikzlibrary{backgrounds}
\usetikzlibrary{patterns}
\usepackage{xfrac} 
\usepackage{nicefrac}  
\usepackage{upgreek} 
\usepackage{cases} 
\usepackage{subcaption} 
\usepackage{siunitx}
\usepackage{empheq}
\usepackage{xcolor}

\DeclareMathOperator {\diag}{diag}
\DeclareMathOperator {\Diag}{Diag}
\DeclareMathOperator {\Diagg}{Diagg}
\DeclareMathOperator {\tr}{tr}
\DeclareMathOperator {\st}{s.t.}
\DeclareMathOperator {\Exp}{E}

\DeclareMathAlphabet{\pazocal}{OMS}{zplm}{m}{n}

\def\a{\mathbf{a}}

\def\x{\mathbf{x}}

\def\A{\mathbf{A}}

\def\h{\mathbf{h}}
\def\H{\mathbf{H}}

\def\0{{\mathbf{0}}}
\def\A{\mathbf{A}}

\def\P{\mathbf{P}}

\def\bPsi{\boldsymbol{\Psi}}
\def\bpsi{\boldsymbol{\psi}}
\def\bPhi{\boldsymbol{\Phi}}
\def\bSigma{\boldsymbol{\Sigma}}
\def\bOmega{\boldsymbol{\Omega}}
\def\bphi{\boldsymbol{\upvarphi}}

\def\I{\mathbf{I}}
\def\bGamma{\mathbf{\Gamma}}
\def\bLambda{\mathbf{\Lambda}}

\renewcommand{\-}{\hskip 0.13em{-}\hskip 0.13em}
\newcommand*{\transp}{{\scriptscriptstyle{T}}}
\newcommand*{\herm}{{\scriptscriptstyle{H}}}
\newcommand*{\inv}{{\scriptscriptstyle{-1}}}

\allowdisplaybreaks
\graphicspath{{./fig/}}

\pgfdeclarelayer{bg}    	
\pgfsetlayers{bg,main}  
\def\refConv{
{\protect\tikz \protect\draw[line width=0.7pt, fill=black!10!red, draw=none] rectangle (.6em, .6em);}
}
\def\refFactNoUp{
{\protect\tikz \protect\draw[preaction={line width=0.7pt, fill=black!10!blue}, draw=none, pattern=north west lines, pattern color=white!60!blue] rectangle (.6em, .6em);}
}
\def\refFactUp{
{\protect\tikz \protect\draw[preaction={line width=0.7pt, fill=black!10!blue}, draw=none, pattern=north east lines, pattern color=white!60!blue] rectangle (.6em, .6em);}
}
\newlength\fheight 			
\newlength\fwidth 

\title{Square Root-based Multi-Source Early PSD Estimation and Recursive RETF Update in Reverberant Environments by Means of the Orthogonal Procrustes Problem}
       
\author{Thomas~Dietzen,~
        Simon~Doclo,~\IEEEmembership{Senior Member,}
        Marc~Moonen,~\IEEEmembership{Fellow,}
        and~Toon~van~Waterschoot~\IEEEmembership{Member} 
              
\thanks{T. Dietzen and T. van Waterschoot are with KU Leuven, Dept. of Electrical Engineering (ESAT), STADIUS Center for Dynamical Systems, Signal Processing and Data Analytics and ETC Technology Cluster Electrical Engineering, Leuven, Belgium. S. Doclo is with University of Oldenburg, Dept. of Medical Physics and Acoustics and the Cluster of Excellence Hearing4all, Oldenburg, Germany. M. Moonen is with KU Leuven, ESAT, STADIUS.}}       

\begin{document}

\maketitle

\begin{abstract}
Multi-channel short-time Fourier transform (STFT) domain-based processing of reverberant microphone signals commonly relies on power-spectral-density (PSD) estimates of early source images, where early refers to reflections contained within the same STFT frame.
State-of-the-art approaches to multi-source early PSD estimation, given an estimate of the associated relative early transfer functions (RETFs), conventionally minimize the approximation error defined with respect to the early correlation matrix, requiring non-negative inequality constraints on the PSDs. Instead, we here propose to factorize the early correlation matrix and minimize the approximation error defined with respect to the early-correlation-matrix square root. The proposed minimization problem -- constituting a generalization of the so-called orthogonal Procrustes problem -- seeks a unitary matrix and the square roots of the early PSDs up to an arbitrary complex argument, making non-negative inequality constraints redundant. A solution is obtained iteratively, requiring one singular value decomposition (SVD) per iteration. The estimated unitary matrix and early PSD square roots further allow to recursively update the RETF estimate, which is not inherently possible in the conventional approach. 
An estimate of the said early-correlation-matrix square root itself is obtained by means of the generalized eigenvalue decomposition (GEVD), where we further propose to restore non-stationarities by desmoothing the generalized eigenvalues in order to compensate for inevitable recursive averaging.
Simulation results indicate fast convergence of the proposed multi-source early PSD estimation approach in only one iteration if initialized appropriately, and better performance as compared to the conventional approach. 
\end{abstract}

\section{Introduction}
\label{sec:intro}

\IEEEPARstart{I}{n} many multi-microphone signal processing applications, the recorded microphone signals constitute a mixture of several spatially diverse components,  originating from different sources, bearing reverberation and noise.
As far as speech is concerned, one typically admits early reflections, while late reverberant components deteriorate the perceived quality and intelligibility  \cite{beutelmann06}.
In order to process the various mixture components, many techniques heavily rely on estimates of their power spectral densities (PSDs) \cite{loizou2007speech, benesty2008microphone, gannot2017consolidated}.

In recent years, a number of multi-microphone approaches to the estimation of early speech PSDs, late reverberant PSDs, and/or noise PSDs have been proposed, which rely on a spatial correlation matrix model in the short-time Fourier transform (STFT) domain \cite{quang04, Thiergart2013AiMf, BraunH15, SchwartzGH16, Huang16, KuklasinskiDJJ16, SchwartzGH16b,  KodrasiIna2018JLRa, Braun18TASLP, KodrasiD18, Koutrouvelis18}.
In order to estimate these PSDs, some parameters of the correlation matrix model are assumed to be known or estimated beforehand, such as the direction(s) of arrival (DoA(s)) or the relative early transfer function(s) (RETF(s)) associated to the source(s) \cite{quang04, Thiergart2013AiMf, BraunH15, SchwartzGH16, Huang16, KuklasinskiDJJ16, SchwartzGH16b, KodrasiIna2018JLRa, Braun18TASLP}, or the spatial coherence matrix of the noise or the late reverberant component, where in particular the latter is commonly modeled as a spatially diffuse sound field \cite{Thiergart2013AiMf, BraunH15, SchwartzGH16, KuklasinskiDJJ16, SchwartzGH16b, KodrasiIna2018JLRa, Braun18TASLP, KodrasiD18, Koutrouvelis18, jacobsen2000coherence}.
It should be noted that majority of these approaches consider a single source \cite{quang04, SchwartzGH16, KuklasinskiDJJ16, SchwartzGH16b, KodrasiIna2018JLRa, Braun18TASLP,  KodrasiD18}, while only some consider multiple sources \cite{Thiergart2013AiMf, BraunH15, Huang16, Koutrouvelis18}, which is the focus of this paper.

In \cite{SchwartzGH16, KuklasinskiDJJ16}, the early speech and late reverberant PSD estimates are obtained by maximum-likelihood estimation, where in \cite{SchwartzGH16}, both are estimated jointly, and in \cite{KuklasinskiDJJ16}, the late reverberant PSD estimation relies on blocking the early speech component.
Given particular coherence matrix estimates, e.g., defined from DoA or RETF estimates (for point sources) or assumptions on the spatial nature of the sound field (for noise and late reverberation), other estimators rely on Frobenius-norm minimization of the approximation error defined with respect to an estimate of the associated correlation matrix component \cite{quang04, Thiergart2013AiMf, BraunH15, SchwartzGH16b, Huang16, KodrasiIna2018JLRa, Koutrouvelis18}.
Specifically, in \cite{quang04}, the speech PSD is estimated by minimizing the approximation error defined with respect to an estimate of the speech-only correlation matrix component (while reverberation is not considered).
In a similar manner, in \cite{Thiergart2013AiMf}, considering multiple sources, the early PSDs are estimated from an estimate of the early correlation matrix component.
In \cite{BraunH15},  the late reverberant PSD is estimated from an estimate of a blocking-based correlation matrix, generated by blocking the direct components, while the multiple early PSDs are estimated as in \cite{Thiergart2013AiMf}. 
Likewise, one may also jointly estimate several PSDs associated to different kinds of coherence matrices, e.g., one may jointly estimate early speech PSD(s), the late reverberant PSD, and noise PSD(s) \cite{SchwartzGH16b, Huang16, KodrasiIna2018JLRa}. 
In \cite{Koutrouvelis18}, joint estimation of the RETFs, the early speech PSDs, the late reverberant PSD, and the noise PSDs is proposed using simultaneous confirmatory factor analysis in multiple frames, i.e. by jointly minimizing a number of approximation errors defined over several frames, during which the RETFs are assumed to be stationary.
Note that the PSD estimates based on this type of minimization problem are not inherently guaranteed to be non-negative, requiring either non-negative thresholding, or, alternatively, non-negative inequality constraints.
In \cite{KodrasiD18}, the estimation of the late reverberant PSD is based on a subspace decomposition, outperforming the late reverberant PSD estimators in \cite{BraunH15, SchwartzGH16, KuklasinskiDJJ16}, while the early speech PSD estimate is obtained from the decision-directed approach \cite{ephraim84}. 

In this contribution, we are mainly concerned with early PSD estimation and recursive RETF updates for multiple sources in reverberant environments, given initial estimates of the associated RETFs.
Instead of minimizing the approximation error defined with respect to an estimate of the early correlation matrix, however, we propose to factorize the early correlation matrix and minimize the approximation error defined with respect to the early-correlation-matrix square root.
Instead of directly estimating the early PSDs, the proposed minimization problem seeks a unitary matrix and the square roots of the early PSDs up to an arbitrary complex argument, making non-negative thresholding or non-negative inequality constraints redundant. 
The proposed minimization problem constitutes a generalization \cite{everson98} of the so-called orthogonal Procrustes problem \cite{schonemann1966generalized, MantonJH2002} and may be solved iteratively, requiring one singular value decomposition (SVD) per iteration.
The estimated unitary matrix and early PSD square roots further allow us to recursively update the RETF estimate, which is not inherently possible in the conventional approach.
An estimate of the said early-correlation-matrix square root itself is obtained from an estimate of the microphone signal correlation matrix by means of the generalized eigenvalue decomposition (GEVD). 
Hereat, in order to compensate for the inevitable recursive averaging in the microphone-signal-correlation-matrix estimation, we further propose to restore non-stationarities by desmoothing the generalized eigenvalues.
Simulation results indicate fast convergence of the proposed multi-source early PSD estimation approach in only one iteration if initialized appropriately, and better performance as compared to the conventional approach in terms of the relative squared PSD estimation error and the signal-to-interference ratio \cite{VincentGF06} measuring the source-component separation.
A MATLAB implementation is available at \cite{taslp19aCode}.

The remainder of this paper is organized as follows. 
In Sec. \ref{sec:sm}, we introduce the signal model.
Given an estimate of the early correlation matrix component, some state-of-the-art approaches to early PSD estimation are reviewed in Sec. \ref{sec:sota}, while the proposed approach, given an estimate of the early-correlation-matrix square root, is presented in Sec. \ref{sec:proposed}.
In Sec. \ref{sec:GEVD}, we discuss the estimation of the required early correlation matrix component and its factorization.
The proposed approach is evaluated in Sec. \ref{sec:sim}, followed by a conclusion in Sec. \ref{sec:conclusion}.

\section{Signal Model}
\label{sec:sm}

Throughout the paper, we use the following notation: vectors are denoted by lower-case boldface letters, matrices by upper-case boldface letters, $\mathbf{I}$ and $\mathbf{0}$ denote identity and zero matrices, $\mathbf{i}$ and $\mathbf{1}$ denote the first column of $\mathbf{I}$ and a vector of ones, respectively, $\mathbf{A}^\transp$, $\mathbf{A}^\herm$, and $\Exp[\mathbf{A}]$ denote the transpose, the complex conjugate transpose or Hermitian, and the expected value of a matrix $\mathbf{A}$.
The operation $\diag[\A]$ creates a column vector from the diagonal elements of a square matrix $\A$, $\Diag[\a]$ and $\Diag[\a^\transp]$ create a diagonal matrix with the elements of $\a$ on its diagonal,  $\Diagg[\A] = \Diag\bigl[\diag[\A]\bigr]$ zeros the off-diagonal elements of $\A$, and $\tr[\A]$ denotes the trace of $\A$.
For non-negative $\a \in \mathbb{R}^N$, $\a^{\nicefrac{1}{2}} \in \mathbb{C}^N$ denotes a complex vector with arbitrary complex argument that satisfies $\Diag[\a^{\nicefrac{\herm}{2}}]\a^{\nicefrac{1}{2}} = \a$, and hence $\bigl\vert\a^{\nicefrac{1}{2}}\bigr\vert = \sqrt{\a}$, with absolute value and  non-negative square-root applied element-wise.
The operation $\max[\a_1,\a_2]$ returns a vector of the element-wise maxima of $\a_1$ and $\a_2$.
$\Vert\A\Vert_\textsl{F}$ denotes the Frobenius norm of $\A$, whereas $\Vert\a\Vert_2$ denotes the Euclidian norm of $\a$.
Row $i$ and column $j$ of $\A$ are denoted as $[\A]_{i,:}$ and $[\A]_{:,j}$, respectively, the element at their intersection as  $[\A]_{i,j}$, 
and submatrices spanning rows $i_1$ to $i_2$ or columns $j_1$ to $j_2$ as $[\A]_{i_1:i_2,:}$ and $[\A]_{:,j_1:j_2}$, respectively. 
 $\Re[a]$ and  $\Im[a]$ denote the real and imaginary part of $a \in \mathbb{C}$.

In the STFT domain, with $l$ and $k$ indexing the frame and  the frequency bin, respectively, let $x_m(l,k)$ with $m=1,\dots,M$ denote the $m^\text{th}$ microphone signal, with $M$ the number of microphones.
In the following, we treat all frequency bins independently and hence omit the frequency index.
We define the stacked microphone signal vector $\x(l) \in \mathbb{C}^{M}$,
\begin{align}
\x(l) &= 
\begin{pmatrix}
x_1(l) & \cdots & x_{M}(l)
\end{pmatrix}^{\transp} 
\label{eq:sm:y_stacked}
\end{align} 
composed of the reverberant signal components $\x_n(l)$ with $n=1,\dots,N$ originating from $N$ point sources, defined equivalently to (\ref{eq:sm:y_stacked}), i.e.
\begin{align}
{\x}(l) &= \sum_{n=1}^N{\x}_n(l). 
\label{eq:sm:y_decomp}
\end{align}
Each reverberant signal component $\x_n(l)$ may be decomposed into the early and late reverberant component ${\x}_{n|\textsl{e}}(l)$ and ${\x}_{n|\ell}(l)$, i.e. 
\begin{align}
{\x}_n(l) &= {\x}_{n|\textsl{e}}(l)+ {\x}_{n|\ell}(l),
\label{eq:sm:xn_decomp}
\end{align}
which are commonly parted by the arrival time of the therein contained reflections and assumed to have distinct spatial properties as outlined below. 
Early reflections are assumed to arrive within the same frame,
with the early components in ${\x}_{n|\textsl{e}}(l)$ related by the RETF in $\h_n(l) \in \mathbb{C}^{M}$, i.e.
\begin{align}
{\x}_{n|\textsl{e}}(l) &=  \h_n(l)s_{n}(l). 
\label{eq:sm:xne_RETF}
\end{align}
Here, without loss of generality, the RETF $\h_n(l)$ is assumed to be relative to the first microphone, i.e. $\mathbf{i}^{\transp}\h_n(l) = [\h_n(l)]_1 = 1$, and $s_{n}(l) = [{\x}_{n|\textsl{e}}(l)]_1$ denotes the early component in the first microphone originating from the  $n^{\text{th}}$ source, in the following referred to as early source image. 
We define the stacked RETF matrix ${\H}(l) \in \mathbb{C}^{M \times N}$, yielding
\begin{align}
 {\H}(l) & = 
 \begin{pmatrix}
 \h_1(l) & \cdots &\h_N(l)
 \end{pmatrix},
 \label{eq:sm:H_stacked}\\
  \mathbf{i}^{\transp}{\H}(l) &= [{\H}(l)]_{1,:\,} = \mathbf{1}^{\transp}.
  \label{eq:sm:H_constr}
\end{align}
Similarly, we stack $s_{n}(l)$ into $\mathbf{s}(l) \in \mathbb{C}^N$, i.e.  
\begin{align}  
\mathbf{s}(l) &= 
  \begin{pmatrix}
  s_{1}(l) &\cdots &s_{N}(l)
  \end{pmatrix}^\transp, 
  \label{eq:sm:xne_stacked}
\end{align}
such that the sum of the early components ${\x}_{n|\textsl{e}}(l)$ may be expressed more compactly as
\begin{align}
 \sum_{n=1}^N{\x}_{n|\textsl{e}}(l) = {\H}(l)\mathbf{s}(l). 
 \label{eq:sm:sumxne_matmult}
\end{align}
Further, we assume that ${\x}_{n|\textsl{e}}(l)$ and ${\x}_{n|\ell}(l)$ are mutually uncorrelated {within} frame $l$.
Let $\bPsi_{x}(l) = \Exp[\x(l)\x^\herm(l)] \in \mathbb{C}^{M \times M}$ denote the microphone signal correlation matrix, and let the early and late reverberant correlation matrix ${\bPsi}_{x_\textsl{e}}(l)$ and ${\bPsi}_{x_\ell}(l)$ be similarly defined.
With (\ref{eq:sm:xn_decomp})--(\ref{eq:sm:sumxne_matmult}), we then find
\begin{align}
{\bPsi}_{x}(l)  = {\bPsi}_{x_\textsl{e}}(l) + {\bPsi}_{x_\ell}(l), 
\label{eq:sm:Psiy}
\end{align}
wherein ${\bPsi}_{x_\textsl{e}}(l)$ generally has rank $N$ and is expressed by 
\begin{align}
\Aboxed{
{\bPsi}_{x_\textsl{e}}(l)  &= {\H}(l){\bPhi}_{s}(l){\H}^\herm(l),}
\label{eq:sm:Psixe}\\
{\bPhi}_{s}(l) &= \Diag[ {\bphi}_{s}(l)], 
\label{eq:sm:Phixe}\\
 {\bphi}_{s}(l) &=
 \begin{pmatrix}
 \varphi_{s_{1}}(l) & \cdots & \varphi_{s_{N}}(l)
 \end{pmatrix}^{\transp},
 \label{eq:sm:phixe}
  \end{align}
 with $\varphi_{s_{n}}(l)$ denoting the PSD of the early source image $s_n(l)$.
 Note that applying (\ref{eq:sm:H_constr}) to (\ref{eq:sm:Psixe})--(\ref{eq:sm:Phixe}) while using $\mathbf{1}^{\transp}{\bPhi}_{s}(l)\mathbf{1} = \mathbf{1}^{\transp}{\bphi}_{s}(l)$, we find that
\begin{align}
\Aboxed{
\mathbf{i}^{\transp}{{\bPsi}}{}_{x_\textsl{e}}(l)\mathbf{i} = [{{\bPsi}}{}_{x_\textsl{e}}(l)]_{1,1} = \mathbf{1}^{\transp}{\bphi}_{s}(l),} \label{eq:sm:phixe_constr}
\end{align}
i.e. the sum of the early PSDs $\varphi_{s_{n}}(l)$ equals $[{{\bPsi}}{}_{x_\textsl{e}}(l)]_{1,1}$.
 Assuming that ${\x}_{n|\ell}(l)$ may be modeled as diffuse \cite{Thiergart2013AiMf, BraunH15, SchwartzGH16, KuklasinskiDJJ16, SchwartzGH16b, KodrasiIna2018JLRa, Braun18TASLP, KodrasiD18,  Koutrouvelis18, jacobsen2000coherence} with coherence matrix $\mathbf{\Gamma}  \in \mathbb{C}^{M \times M}$, which may be computed from the microphone array geometry \cite{jacobsen2000coherence} and is therefore considered to be known in the remainder, we may write ${\bPsi}_{x_\ell}(l)$ as
 \begin{align}
 {\bPsi}_{x_\ell}(l) &= {\varphi}_{x_\ell}(l)\mathbf{\Gamma},  \label{eq:sm:Psixl}\\
  \text{with}\quad{\varphi}_{x_\ell}(l) &= \sum_{n=1}^N \varphi_{x_{n|\ell}}(l),
  \label{eq:sm:phid}
 \end{align}
and $\varphi_{x_{n|\ell}}(l)$ denoting the PSD of the  late reverberant component ${\x}_{n|\ell}(l)$.
The PSDs  ${\bphi}_{s}(l)$ and ${\varphi}_{x_\ell}(l)$ may be highly non-stationary,  especially if the point sources are speech sources, while the associated coherence matrices $\h_n(l)\h_n^\herm(l)$ and  $\mathbf{\Gamma}$ are commonly assumed to be comparably slowly time-varying or even time-invariant.
 
Note that with (\ref{eq:sm:Psixl})--(\ref{eq:sm:phid}), one may easily include further diffuse components, e.g., babble noise, without formally changing the signal model.
However, since in this paper, we are mainly concerned with the estimation of the early PSDs ${\bphi}_{s}(l)$ and the recursive updating of the estimate of the RETFs ${\H}(l)$, we restrict the discussion and simulations, cf. Sec. \ref{sec:sim}, to the example of late reverberation for the sake of conciseness.

Further, note that while the above signal model is commonly and effectively used \cite{Thiergart2013AiMf, BraunH15, SchwartzGH16, KuklasinskiDJJ16, SchwartzGH16b,  KodrasiIna2018JLRa, Braun18TASLP, KodrasiD18, Koutrouvelis18} due to its simplicity, it may be said to be deficient in a number of aspects.
The assumption that ${\x}_{n|\textsl{e}}(l)$ and ${\x}_{n|\ell}(l)$ are mutually uncorrelated {within} frame $l$ may be violated due to overlapping windows in the STFT-processing or source signals remaining correlated over several frames.
The assumption that ${\bPsi}_{x_\textsl{e}}(l)$ in (\ref{eq:sm:Psixe}) has rank $N$ implicitly relies on the assumption that the frequency bins may be treated independently, ignoring cross-bin dependencies \cite{avergel07}.
Finally, related to that, there may be components that may be modeled neither by the rank-$N$ component ${\bPsi}_{x_\textsl{e}}(l)$  in (\ref{eq:sm:Psixe}) nor by the diffuse component ${\bPsi}_{x_\ell}(l)$  in (\ref{eq:sm:Psixl}), depending on the geometry and physical properties of the acoustic environment.
 
In the remainder, as we mostly consider the single frame $l$ only, we also drop the frame index for conciseness and refer back to it only where necessary, namely when we differentiate the frames $l$ and $l-1$ in recursive equations.

\section{Early PSD Estimation based on the Early Correlation Matrix}
\label{sec:sota}

In this section, we discuss some state-of-the-art approaches to the estimation of the early PSDs ${\bphi}_{s}$ based on the signal model in (\ref{eq:sm:Psixe})--(\ref{eq:sm:phixe_constr}).
In the following, we refer to (\ref{eq:sm:Psixe})--(\ref{eq:sm:phixe_constr}) as the {conventional} signal model.
We develop our discussion from the premise that estimates $\hat{{\bPsi}}_{x_\textsl{e}}$ and $\hat{\H}$ 
of the early correlation matrix ${{\bPsi}}_{x_\textsl{e}}$ and the RETFs ${\H}$ in (\ref{eq:sm:Psixe}) are readily available.
Throughout the paper, despite being irrelevant to the approaches discussed in this section, we consider  $\hat{{\bPsi}}_{x_\textsl{e}}$ to generally have rank $N$, similar to ${{\bPsi}}_{x_\textsl{e}}$.
A rank-$N$ estimator of ${{\bPsi}}_{x_\textsl{e}}$ is described in Sec. \ref{sec:GEVD}.
Further, we assume that   $\hat{\H}$ satisfies $\mathbf{i}^{\transp}\hat{\H} = \mathbf{1}^{\transp}$, cf.  ${\H}$ in (\ref{eq:sm:H_constr}).
 
Given the estimates of a early correlation matrix $\hat{{\bPsi}}{}_{x_\textsl{e}}$ and the therein superimposed coherence matrices $\hat{\h}_n\hat{\h}{}^\herm_n$, one may estimate the associated PSDs $\varphi_{s_{n}}$, cf. (\ref{eq:sm:H_stacked}), (\ref{eq:sm:Psixe})--(\ref{eq:sm:Phixe}), as described in \cite{Thiergart2013AiMf, SchwartzGH16b, Huang16, KodrasiIna2018JLRa}\footnote{In \cite{Thiergart2013AiMf}, as in our case, point-source coherence matrices of rank one are considered, while  in \cite{SchwartzGH16b, Huang16, KodrasiIna2018JLRa}, without rendering a difference in the principle approach, general coherence matrices are considered.}.
Adopting this approach, 
we define the approximation error as a function of ${\bphi}_{s}$ as
\begin{align}
\mathbf{E}_{\textsl{c}}({\bphi}_{s}) &= \hat{{\bPsi}}{}_{x_\textsl{e}} -  \hat{\H}\Diag[{\bphi}_{s}]\hat{\H}^\herm, \label{eq:sota:Ec}
\end{align}
where the subscript $\textsl{c}$ stands for conventional.
The early PSDs ${\bphi}_{s}$ can then be estimated by Frobenius-norm minimization of the approximation error followed by non-negative thresholding, i.e.
  \begin{align}
	\hat{\bphi}'_{s} &= \underset{{\bphi}_{s}}{\arg\min}
	\,\bigl\Vert\mathbf{E}_{\textsl{c}}({\bphi}_{s}) \bigr\Vert_\textsl{F}^2
	\label{eq:sota:phixeHatPrime_minprob},\\
	\hat{\bphi}_{s} &= \text{max}[\hat{\bphi}'_{s},\,\mathbf{0}].
	\label{eq:sota:phixeHat}
\end{align}
The non-negative thresholding in (\ref{eq:sota:phixeHat}) is necessary as the elements of $\hat{\bphi}'_{s}$ in (\ref{eq:sota:phixeHatPrime_minprob}) may in fact be negative, conflicting with the notion of ${\bphi}_{s}$ being a vector of PSDs.
If $\hat{\H}^\herm\hat{\H}$ has full rank, which (without sufficiency) requires $N \leq M$, the problem in (\ref{eq:sota:phixeHatPrime_minprob}) has a unique solution given by
 \begin{align}
	\hat{\bphi}'_{s} &= {\mathbf{A}}{}_{\textsl{c}_0}^{\inv}{\mathbf{b}}_{\textsl{c}_0},\label{eq:sota:phixeHatPrime_sol}
\end{align}
where ${\mathbf{A}}{}_{\textsl{c}_0} \in \mathbb{R}^{N\times N}$  and ${\mathbf{b}}_{\textsl{c}_0} \in \mathbb{R}^{N}$ are defined by
 \begin{align}
	[{\mathbf{A}}_{\textsl{c}_0}]_{n,n'} &= \vert\hat{\h}{}^\herm_n\hat{\h}_{n'}\vert^{2},\label{eq:sota:Actilde}\\
	{\mathbf{b}}_{\textsl{c}_0} &= \diag[\hat{\H}^\herm\hat{{\bPsi}}{}_{x_\textsl{e}}\hat{\H}].\label{eq:sota:bctilde}
\end{align}
Alternatively, instead of simple thresholding after solving (\ref{eq:sota:phixeHatPrime_minprob}), one may solve the minimization problem subject to the non-negative inequality constraint ${\bphi}_{s} \geq \mathbf{0}$, as proposed in \cite{Koutrouvelis18}.
In addition to this, one may further impose a soft constraint on $\mathbf{1}^{\transp}{\bphi}_{s}$ corresponding to (\ref{eq:sm:phixe_constr}), i.e. one may define a soft-constraint error as a function of ${\bphi}_{s}$ to be penalized as
\begin{align}
e_{\textsl{c}}({\bphi}_{s}) &= [\hat{{\bPsi}}{}_{x_\textsl{e}}]_{1,1}-\mathbf{1}^{\transp}{\bphi}_{s}\nonumber\\&= [\mathbf{E}_{\textsl{c}}({\bphi}_{s})]_{1,1}.\label{eq:sota:ec}
\end{align}
The resulting minimization problem can then be written as 
\begin{empheq}[box=\fbox]{align}
	\hat{\bphi}_{s} &= \underset{{\bphi}_{s}}{\arg\min}
	\,\bigl\Vert\mathbf{E}_{\textsl{c}}({\bphi}_{s}) \bigr\Vert_\textsl{F}^2+\alpha \bigl\vert e_{\textsl{c}}({\bphi}_{s})\bigl\vert^2\nonumber\\[-2pt]
&\hphantom{=\arg\,\,}\st  \quad{\bphi}_{s} \geq \mathbf{0}, \label{eq:sota:phixeHat_minprob}
\end{empheq}
where $\alpha$ is the penalty factor.
For $\alpha \rightarrow \infty$, a hard constraint $\mathbf{1}^{\transp}{\bphi}_{s} = [\hat{{\bPsi}}{}_{x_\textsl{e}}]_{1,1}$ is introduced, which may however not be desirable due to potential estimation errors in $[\hat{{\bPsi}}{}_{x_\textsl{e}}]_{1,1}$.
Note that in \cite{Koutrouvelis18},  instead of a soft constraint, a box constraint on $\mathbf{1}^{\transp}{\bphi}_{s}$ has been used. 
For the sake of comparison to the algorithm proposed in Sec. \ref{sec:proposed}, however, we restrict our discussion to the soft constraint.
Due to the inequality constraint ${\bphi}_{s} \geq \mathbf{0}$, the problem in (\ref{eq:sota:phixeHat_minprob}) does not have a closed-form solution and may require several iterations in order to be solved.
Using the proximal gradient method \cite{parikh2014proximal, antonello2018proximal}, we may solve (\ref{eq:sota:phixeHat_minprob}) by iterating the below set of equations until convergence is reached,
 \begin{align}
	\hat{\bphi}{}'^{(i)}_{s} &= \hat{\bphi}{}^{(i-1)}_{s} + \mu\bigl(\mathbf{b}_{\textsl{c}}-\mathbf{A}_{\textsl{c}}\hat{\bphi}{}^{(i-1)}_{s}\bigr),\label{eq:sota:proxgradgrad}\\
	\hat{\bphi}{}^{(i)}_{s} &= \max[\hat{\bphi}{}'^{(i)}_{s},\mathbf{0}], \label{eq:sota:proxgradthres}
\end{align}
where $i$ is the iteration index, $\mu$ the step-size, and $\mathbf{b}_{\textsl{c}}-\mathbf{A}_{\textsl{c}}\hat{\bphi}{}^{(i-1)}_{s}$ the gradient with ${\mathbf{A}}{}_{\textsl{c}} \in \mathbb{R}^{N\times N}$ and ${\mathbf{b}}_{\textsl{c}} \in \mathbb{R}^{N}$ defined by
 \begin{align}
	\mathbf{A}_{\textsl{c}} &= {\mathbf{A}}_{\textsl{c}_0} + \alpha\mathbf{1}\mathbf{1}^{\transp},\label{eq:sota:Ac}\\
	\mathbf{b}_{\textsl{c}} &= {\mathbf{b}}_{\textsl{c}_0}  + \alpha[\hat{{\bPsi}}{}_{x_\textsl{e}}]_{1,1}\mathbf{1},\label{eq:sota:bc}
\end{align}
and ${\mathbf{A}}_{\textsl{c}_0}$ and ${\mathbf{b}}_{\textsl{c}_0}$ defined in (\ref{eq:sota:Actilde})--(\ref{eq:sota:bctilde}). 
As initial value, it is straight-forward to choose $\hat{\bphi}{}^{(0)}_{s} = \mathbf{A}_{\textsl{c}}^{\inv}\mathbf{b}_{\textsl{c}}$, which yields the global minimum if  $\hat{\bphi}{}^{(0)}_{s} \geq \mathbf{0}$.
In this case therefore, convergence is reached after one iteration of (\ref{eq:sota:proxgradgrad})--(\ref{eq:sota:proxgradthres}).
In any case, for $\hat{\bphi}{}^{(0)}_{s} = \mathbf{A}_{\textsl{c}}^{\inv}\mathbf{b}_{\textsl{c}}$ and $\alpha = 0$, the estimate obtained after one iteration of (\ref{eq:sota:proxgradgrad})--(\ref{eq:sota:proxgradthres}) corresponds to the estimate  defined by (\ref{eq:sota:phixeHatPrime_minprob})--(\ref{eq:sota:phixeHatPrime_sol}). 
We therefore use (\ref{eq:sota:phixeHat_minprob}) as a reference for comparison in the remainder.
In the following, we refer to (\ref{eq:sota:phixeHat_minprob}) as the conventional minimization problem (conventional MP).

\section{Early PSD Estimation and Recursive RETF Update based on the Early-Correlation- Matrix Square Root}
\label{sec:proposed}

In this section,  in order to estimate the early PSDs ${\bphi}_{s}$, instead of 
defining the approximation error to be minimized with respect to $\hat{{\bPsi}}_{x_\textsl{e}}$ as in (\ref{eq:sota:Ec}), we propose to define the approximation error with respect to the square root $\hat{{\bPsi}}{}^{\nicefrac{1}{2}}_{x_\textsl{e}} \in \mathbb{C}^{M\times N}$ of $\hat{{\bPsi}}{}_{x_\textsl{e}}$, satisfying $\hat{{\bPsi}}_{x_\textsl{e}} = \hat{{\bPsi}}{}^{\nicefrac{1}{2}}_{x_\textsl{e}}\hat{{\bPsi}}{}^{\nicefrac{\herm}{2}}_{x_\textsl{e}}$.
As to be shown, instead of directly estimating the {diagonal} of ${\bPhi}_{s} = \Diag[{\bphi}_{s}]$, the resulting minimization problem now consists in estimating a {unitary} matrix $\bOmega \in \mathbb{C}^{N\times N}$ and the {diagonal} of ${\bPhi}_{s}^{\nicefrac{1}{2}} = \Diag[{\bphi}^{\nicefrac{1}{2}}_{s}]$, which constitutes a generalization \cite{everson98} of the so-called {orthogonal} Procrustes problem \cite{schonemann1966generalized, MantonJH2002}.
Since the early PSDs herein are represented by ${\bphi}_{s} = \Diag[{\bphi}^{\nicefrac{\herm}{2}}_{s}]{\bphi}^{\nicefrac{1}{2}}_{s}$, the corresponding estimate  $\hat{\bphi}_{s}$  is guaranteed to be non-negative, such that a non-negative inequality constraint as in (\ref{eq:sota:phixeHat_minprob}) is not required.
Further, we show that the obtained estimates $\hat{\bOmega}$ and $\hat{\bphi}{}^{\nicefrac{1}{2}}_{s}$ may be used to recursively update the RETF estimate $\hat{\H}$, which is not inherently possible from the estimate $\hat{\bphi}_{s}$ given by (\ref{eq:sota:phixeHat_minprob}).

In Sec. \ref{sec:fact}, as a pre-requisite to our derivation, we discuss the factorization of the conventional signal model in (\ref{eq:sm:Psixe})--(\ref{eq:sm:phixe_constr}), yielding the square-root signal model.
In Sec. \ref{sec:oppPSD}, based upon the square-root signal model and given the estimates $\hat{{\bPsi}}{}^{\nicefrac{1}{2}}_{x_\textsl{e}}$ and  $\hat{\H}$, we then define and solve the square-root minimization problem (square-root MP).
In Sec. \ref{sec:RETFup}, we discuss the recursive updating of the the RETF estimate $\hat{\H}$.

\subsection{Early-Correlation-Matrix Factorization} 
\label{sec:fact}

We consider the factorization of the rank-$N$ matrices on both sides of (\ref{eq:sm:Psixe}).
On the left-hand side of (\ref{eq:sm:Psixe}), we define the square root ${{\bPsi}}{}^{\nicefrac{1}{2}}_{x_\textsl{e}} \in \mathbb{C}^{M\times N}$ such that  ${{\bPsi}}{}^{\nicefrac{1}{2}}_{x_\textsl{e}}{{\bPsi}}{}^{\nicefrac{\herm}{2}}_{x_\textsl{e}} = {{\bPsi}}_{x_\textsl{e}}$. 
Note that the product is invariant to right-multiplication of a particular square root with any  {unitary} matrix, and so ${{\bPsi}}{}^{\nicefrac{1}{2}}_{x_\textsl{e}}$ is not unique.
On the right-hand side of (\ref{eq:sm:Psixe}), with ${\bPhi}_{s}^{\nicefrac{1}{2}} = \Diag[{\bphi}^{\nicefrac{1}{2}}_{s}]$ and $\Diag[{\bphi}^{\nicefrac{\herm}{2}}_{s}]{\bphi}^{\nicefrac{1}{2}}_{s} = {\bphi}_{s}$, we define the square root $\H{\bPhi}_{s}^{\nicefrac{1}{2}}$  such that 
 $\H{\bPhi}_{s}^{\nicefrac{1}{2}}{\bPhi}_{s}^{\nicefrac{\herm}{2}}\H^\herm = \H{\bPhi}_{s}\H^\herm$.
Note that  while the magnitude of the elements in ${\bphi}^{\nicefrac{1}{2}}_{s} \in \mathbb{C}^N$ is well-defined, namely by $\bigl\vert{\bphi}^{\nicefrac{1}{2}}_{s}\bigr\vert = \sqrt{{\bphi}_{s}}$, their complex argument may be chosen arbitrarily, and so ${\bPhi}_{s}^{\nicefrac{1}{2}}$ is not unique.
The non-uniqueness of both square roots implies that while their respective products on both sides of (\ref{eq:sm:Psixe}) coincide, the said square roots themselves generally do not, i.e. we have ${{\bPsi}}{}^{\nicefrac{1}{2}}_{x_\textsl{e}} \neq \H{\bPhi}_{s}^{\nicefrac{1}{2}}$.
Hence, for a particular ${{\bPsi}}{}^{\nicefrac{1}{2}}_{x_\textsl{e}}$ and ${\bPhi}_{s}^{\nicefrac{\herm}{2}}$,
we introduce the {unitary} matrix $\bOmega \in \mathbb{C}^{N \times N}$, which is such that 
 ${{\bPsi}}{}^{\nicefrac{1}{2}}_{x_\textsl{e}}\bOmega$ and $\H{\bPhi}_{s}^{\nicefrac{1}{2}}$ do coincide, i.e. we may summarize
\begin{align}
\Aboxed{
{{\bPsi}}{}^{\nicefrac{1}{2}}_{x_\textsl{e}}\bOmega &= \H{\bPhi}_{s}^{\nicefrac{1}{2}},}
\label{eq:fact:Psixe_fact}\\
{\bPhi}_{s}^{\nicefrac{1}{2}} &= \Diag[{\bphi}^{\nicefrac{1}{2}}_{s}],
 \label{eq:fact:Sigmaxe_decomp}\\
\bOmega\bOmega^\herm &= \I, 
\label{eq:fact:Sigma_orthon}
 \end{align}
where right-multiplying each side of (\ref{eq:fact:Psixe_fact}) with its Hermitian yields (\ref{eq:sm:Psixe}).
At this point, in order to stress the meaning of (\ref{eq:fact:Psixe_fact})--(\ref{eq:fact:Sigma_orthon}), 
we add that  
the column vectors $[{{\bPsi}}{}^{\nicefrac{1}{2}}_{x_\textsl{e}}]_{:,n}$ and $[\H{\bPhi}_{s}^{\nicefrac{1}{2}}]_{:,n} = \h_n{\varphi}_{s}^{\nicefrac{1}{2}}$  form generally {different} bases\footnote{
A particular case is obtained for $N = 1$,
where $\bOmega$ and ${\bPhi}_{s}^{\nicefrac{1}{2}}$ are scalar, while $\H = \h$ and ${{\bPsi}}{}^{{\scriptscriptstyle{{1}/{2}}}}_{x_\textsl{e}} = {\bpsi}{}^{{\scriptscriptstyle{{1}/{2}}}}_{x_\textsl{e}}$ are proportional column vectors.
In this case, given an estimate $\hat{\bpsi}{}^{{\scriptscriptstyle{{1}/{2}}}}_{x_\textsl{e}}$, we may even estimate $\h$ by $\hat{\h} = \hat{{\bpsi}}{}^{{\scriptscriptstyle{{1}/{2}}}}_{x_\textsl{e}}/[\hat{{\bpsi}}{}^{{\scriptscriptstyle{{1}/{2}}}}_{x_\textsl{e}}]_1$, satisfying $[\hat{\h}]_1 = 1$, cf. (\ref{eq:sm:H_constr}).
In essence, despite somewhat different derivation and terminology, this is equivalent to the approach taken in subspace-based single-source RETF estimation \cite{markovichAug09}.
} of the same vector space, and hence 
$\bOmega$ implements a change of basis.
 
Applying (\ref{eq:sm:H_constr}) to (\ref{eq:fact:Psixe_fact})--(\ref{eq:fact:Sigmaxe_decomp}) and noting that 
$\mathbf{1}^{\transp}\Diag[{\bphi}^{\nicefrac{1}{2}}_{s}] = {\bphi}{}^{\nicefrac{\transp}{2}}_{s}$, we find  that ${\bphi}^{\nicefrac{1}{2}}_{s}$ and ${\bOmega}$  satisfy
\begin{align}
\Aboxed{
\mathbf{i}^{\transp}{{\bPsi}}{}^{\nicefrac{1}{2}}_{x_\textsl{e}}{\bOmega} = [{{\bPsi}}{}^{\nicefrac{1}{2}}_{x_\textsl{e}}{\bOmega}]_{1,:} = {\bphi}^{\nicefrac{\transp}{2}}_{s},}\label{eq:fact:Sigma_phixe_constr}
\end{align}
where right-multiplying each side of (\ref{eq:fact:Sigma_phixe_constr}) with its Hermitian yields (\ref{eq:sm:phixe_constr}).
We further note that if ${\bOmega}$ was known for a given square root ${{\bPsi}}{}^{\nicefrac{1}{2}}_{x_\textsl{e}}$, then ${\bphi}^{\nicefrac{1}{2}}_{s}$ could be obtained from (\ref{eq:fact:Sigma_phixe_constr}) immediately.
In the following, we refer to (\ref{eq:fact:Psixe_fact})--(\ref{eq:fact:Sigma_phixe_constr}) as the square-root signal model.

\subsection{Orthogonal Procrustes-based Early PSD Estimate}
\label{sec:oppPSD}

In this section, based on the square-root signal model in (\ref{eq:fact:Psixe_fact})--(\ref{eq:fact:Sigma_phixe_constr}), we seek {unitary} and {diagonal} estimates $\hat{\bOmega}$ and $\Diag[ \hat{\bphi}{}^{\nicefrac{1}{2}}_{s}]$  of  ${\bOmega}$ and $\Diag[{\bphi}{}^{\nicefrac{1}{2}}_{s}]$.
Similarly to Sec. \ref{sec:sota}, we develop our discussion from the premise that estimates $\hat{{\bPsi}}{}^{\nicefrac{1}{2}}_{x_\textsl{e}}$ and $\hat{\H}$ of the early-correlation-matrix square root ${{\bPsi}}{}^{\nicefrac{1}{2}}_{x_\textsl{e}}$ and the RETF ${\H}$ in (\ref{eq:fact:Psixe_fact}) are readily available, 
with $\hat{{\bPsi}}{}^{\nicefrac{1}{2}}_{x_\textsl{e}}$ generally of rank $N$ and $\mathbf{i}^{\transp}\hat{\H} = \mathbf{1}^{\transp}$.
An estimator of ${{\bPsi}}{}^{\nicefrac{1}{2}}_{x_\textsl{e}}$ is described in Sec. \ref{sec:desmoothGEVD}, while Sec. \ref{sec:RETFup} describes a recursive update scheme for $\hat{\H}$.

Similarly to Sec. \ref{sec:sota}, now based on the square-root signal model in (\ref{eq:fact:Psixe_fact})--(\ref{eq:fact:Sigma_orthon})
 instead of the conventional signal model in (\ref{eq:sm:Psixe}), we define the approximation error as a function of ${\bOmega}$ and ${\bphi}^{\nicefrac{1}{2}}_{s}$, i.e.
\begin{align}
\mathbf{E}_{\textsl{sq}}({\bOmega},{\bphi}^{\nicefrac{1}{2}}_{s}) &= \hat{{\bPsi}}{}^{\nicefrac{1}{2}}_{x_\textsl{e}}\bOmega -  \hat{\H}\Diag[ {\bphi}^{\nicefrac{1}{2}}_{s}], 
\label{eq:oppPSD:Ef}
\end{align}
which is akin to $\mathbf{E}_{\textsl{c}}({\bphi}_{s})$ in (\ref{eq:sota:Ec}), and where the subscript $\textsl{sq}$ stands for square root.
Further, now based on the square-root signal model in (\ref{eq:fact:Sigma_phixe_constr})
 instead of the conventional signal model in (\ref{eq:sm:phixe_constr}),
we define a soft-constraint error as a function of ${\bOmega}$ and ${\bphi}^{\nicefrac{1}{2}}_{s}$ to be penalized as
\begin{align}
\mathbf{e}_{\textsl{sq}}({\bOmega},{\bphi}^{\nicefrac{1}{2}}_{s}) &= [\hat{{\bPsi}}{}^{\nicefrac{1}{2}}_{x_\textsl{e}}\bOmega]_{1,:}^{\,\transp} -  {\bphi}^{\nicefrac{1}{2}}_{s} \nonumber\\
&= [\mathbf{E}_{\textsl{sq}}({\bOmega},{\bphi}^{\nicefrac{1}{2}}_{s})]_{1,:}^{\,\transp}, \label{eq:oppPSD:ef}
\end{align}
which is akin to ${e}_{\textsl{c}}({\bphi}_{s})$ in (\ref{eq:sota:ec}).
Note that while ${e}_{\textsl{c}}({\bphi}_{s})$ defines a error on the sum of the early PSDs, $\mathbf{e}_{\textsl{sq}}\bigr({\bOmega},{\bphi}^{\nicefrac{1}{2}}_{s}\bigl)$ instead defines an error on each of the early PSD square roots and is therefore more informative.
Based on (\ref{eq:oppPSD:Ef}), (\ref{eq:oppPSD:ef}), and the {unitary} constraint in (\ref{eq:fact:Sigma_orthon}), 
we define the minimization problem,
\begin{empheq}[box=\fbox]{align}
&\{\hat{\bOmega},\hat{\bphi}^{\nicefrac{1}{2}}_{s}\} = \nonumber\\
&\quad\qquad \underset{\bOmega,\,{\bphi}^{\nicefrac{1}{2}}_{s}}{\arg\min}\,
	\bigl\Vert\mathbf{E}_{\textsl{sq}}({\bOmega},{\bphi}^{\nicefrac{1}{2}}_{s})\bigr\Vert^2_\textsl{F}
	+\alpha \bigl\Vert\mathbf{e}_{\textsl{sq}}({\bOmega},{\bphi}^{\nicefrac{1}{2}}_{s}) \bigr\Vert^2_2 \nonumber\\[-2pt]
&\quad\qquad\hphantom{\arg}\st \quad\bOmega\bOmega^\herm  = \I,
\label{eq:oppPSD:minprob}
\end{empheq}
which is akin to the conventional MP in (\ref{eq:sota:phixeHat_minprob}) and referred to as the square-root minimization problem (square-root MP) in the following.
While the {unitary} constraint in (\ref{eq:oppPSD:minprob}) does not have an equivalent in (\ref{eq:sota:phixeHat_minprob}), the inequality constraint ${\bphi}_{s} \geq \mathbf{0}$  used in (\ref{eq:sota:phixeHat_minprob}) is not required in (\ref{eq:oppPSD:minprob}), as in the square-root signal model, we find that ${\bphi}_{s} = \Diag[{\bphi}^{\nicefrac{\herm}{2}}_{s}]{\bphi}^{\nicefrac{1}{2}}_{s}$, and therefore the corresponding estimate $\hat{\bphi}_{s}$ is guaranteed to be non-negative.
Problems of the kind as in (\ref{eq:oppPSD:minprob}), i.e. Frobenius-norm minimization problems seeking a {unitary} and a {diagonal} matrix, here $\bOmega$ and $\Diag[{\bphi}^{\nicefrac{1}{2}}_{s}]$,
constitute a generalization \cite{everson98} of the so-called {orthogonal} Procrustes problem \cite{everson98, schonemann1966generalized, MantonJH2002}, which seeks a {unitary} matrix only. 
As outlined in the following, under a specific rank condition, the {orthogonal} Procrustes problem has a unique closed-form solution, which is found by means of the SVD \cite{schonemann1966generalized, MantonJH2002}.
The generalized {orthogonal} Procrustes problem, on the contrary, does not have a unique closed-form solution,  but may be solved iteratively \cite{everson98}.
In particular, along the lines of  \cite{everson98}, we propose to solve (\ref{eq:oppPSD:minprob}) by alternatingly \mbox{(re-)}estimating ${\bOmega}$ and ${\bphi}^{\nicefrac{1}{2}}_{s}$ until convergence is reached, namely by solving the {orthogonal} Procrustes problem and the soft-constrained convex problem, respectively,
 \begin{align} 
\hat{\bOmega}{}^{(i)} &= \underset{\bOmega}{\arg\min}\,
	\bigl\Vert\mathbf{E}_{\textsl{sq}}({\bOmega},\hat{\bphi}{}^{\nicefrac{1}{2}|(i-1)}_{s})\bigr\Vert^2_\textsl{F}\nonumber\\[-2pt]
	&\hphantom{= \arg\,\,}\st \quad\bOmega \bOmega^\herm = \I,\label{eq:oppPSD:minprobOmega}
\\[4pt]
\hat{\bphi}{}^{\nicefrac{1}{2}|(i)}_{s} &= \underset{{\bphi}{}^{\nicefrac{1}{2}}_{s}}{\arg\min}\,
	\bigl\Vert\mathbf{E}_{\textsl{sq}}(\hat{\bOmega}{}^{(i)},{\bphi}{}^{\nicefrac{1}{2}}_{s})\bigr\Vert^2_\textsl{F} \nonumber\\[-5pt]
	&\hphantom{= \arg\,\,\st \quad}+ \alpha \bigl\Vert\mathbf{e}_{\textsl{sq}}(\hat{\bOmega}{}^{(i)},{\bphi}{}^{\nicefrac{1}{2}}_{s})\bigr\Vert^2_2,
\label{eq:oppPSD:minprobPhi}
\end{align}
where the soft constraint is applied in (\ref{eq:oppPSD:minprobPhi}) only, i.e. once per iteration. 
Using (\ref{eq:oppPSD:Ef}), by expansion of the Frobenius norm in (\ref{eq:oppPSD:minprobOmega}), it is easily shown \cite{schonemann1966generalized, MantonJH2002} that (\ref{eq:oppPSD:minprobOmega}) is equivalent to 
\begin{align}
\hat{\bOmega}{}^{(i)} &= \underset{\bOmega}{\arg\max}\,\Re\bigl[\tr[\bOmega\mathbf{C}_\textsl{sq}^{(i-1)}]\bigr]\nonumber\\[-2pt]
&\hphantom{= \arg\,}\st \quad\bOmega\bOmega^\herm = \I,\\[3pt]
\text{with} \quad \mathbf{C}^{(i-1)} &= 
\Diag[\hat{\bphi}{}^{\nicefrac{\herm}{2}|(i-1)}_{s}]\hat{\H}^\herm
\hat{{\bPsi}}{}^{\nicefrac{1}{2}}_{x_\textsl{e}}.
\end{align}
If $\mathbf{C}^{(i-1)}$ has full rank,  which (without sufficiency) requires $N \leq M$, the problem in (\ref{eq:oppPSD:minprobOmega}) has a unique closed-form solution, which is based on the SVD of $\mathbf{C}^{\herm|(i-1)}$.
Precisely, if we decompose $\mathbf{C}^{\herm|(i-1)}$ as 
\begin{align}
\mathbf{C}^{\herm|(i-1)} = \mathbf{U}_{\textsl{\tiny L}}\bSigma\mathbf{U}_{\textsl{\tiny R}}^\herm,\label{eq:oppPSD:SVD}
\end{align}
where $\bSigma \in \mathbb{R}^{N\times N}$ is a diagonal matrix of singular values and both $\mathbf{U}_{\textsl{\tiny L}} \in \mathbb{C}^{N\times N}$ and $\mathbf{U}_{\textsl{\tiny R}} \in \mathbb{C}^{N\times N}$ are {unitary}, then $\hat{\bOmega}{}^{(i)}$  is given by
\begin{align}
\hat{\bOmega}{}^{(i)} = \mathbf{U}_{\textsl{\tiny L}}\mathbf{U}_{\textsl{\tiny R}}^\herm. \label{eq:oppPSD:unitSVD}
\end{align}
With (\ref{eq:oppPSD:Ef}) and (\ref{eq:oppPSD:ef}), the solution to (\ref{eq:oppPSD:minprobPhi}) is easily found as 
\begin{align}
\hat{\bphi}{}^{\nicefrac{1}{2}|(i)}_{s} &=  \mathbf{A}_{\textsl{sq}}^{\inv}\mathbf{b}^{(i)}_{\textsl{sq}}, \label{eq:oppPSD:solPhi}
\end{align}
with ${\mathbf{A}}{}_{\textsl{sq}} \in \mathbb{R}^{N\times N}$ and ${\mathbf{b}}^{(i)}_{\textsl{sq}} \in \mathbb{C}^{N}$ are defined by
\begin{align}
\mathbf{A}_{\textsl{sq}} &= \Diagg[\hat{\H}^\herm\hat{\H}] + \alpha\mathbf{I},\label{eq:oppPSD:Af}\\
		\mathbf{b}^{(i)}_{\textsl{sq}} &= \diag[\hat{\H}^\herm(\mathbf{I}+\alpha\mathbf{i}\mathbf{i}^{\transp})\hat{{\bPsi}}{}^{\nicefrac{1}{2}}_{x_\textsl{e}}\hat{\bOmega}{}^{(i)}]\nonumber\\
		&=\diag[\hat{\H}^\herm\hat{{\bPsi}}{}^{\nicefrac{1}{2}}_{x_\textsl{e}}\hat{\bOmega}{}^{(i)}] + \alpha[\hat{\bPsi}{}^{\nicefrac{1}{2}}_{x_\textsl{e}}\hat{\bOmega}{}^{(i)}]_{1,:}^{\,\transp}. \label{eq:oppPSD:bf}
\end{align}
The set of equations (\ref{eq:oppPSD:Af})--(\ref{eq:oppPSD:bf}) is akin to  (\ref{eq:sota:Ac})--(\ref{eq:sota:bc}) for the conventional MP. 
Note that for $\alpha \rightarrow \infty$, the soft constraint in the square-root MP in (\ref{eq:oppPSD:minprobPhi}) becomes a hard constraint and, moreover, solely determines $\hat{\bphi}{}^{\nicefrac{1}{2}|(i)}_{s}$, namely as $\hat{\bphi}{}^{\nicefrac{1}{2}|(i)}_{s} = [\hat{\bPsi}{}^{\nicefrac{1}{2}}_{x_\textsl{e}}\hat{\bOmega}{}^{(i)}]_{1,:}^{\,\transp}$ according to (\ref{eq:oppPSD:solPhi})--(\ref{eq:oppPSD:bf}).
This is not the case for the soft constraint in the conventional MP in (\ref{eq:sota:phixeHat_minprob}).
 
Note that since the problem in (\ref{eq:oppPSD:minprob}) is non-convex, the iteration in (\ref{eq:oppPSD:minprobOmega})--(\ref{eq:oppPSD:minprobPhi}) is not guaranteed to converge to a global minimum \cite{everson98}.
The initial value $\hat{\bphi}{}^{\nicefrac{1}{2}|(0)}_{s}$ of the iteration may, e.g., be chosen based on the sum constraint in (\ref{eq:sm:phixe_constr}) as $\hat{\bphi}{}^{\nicefrac{1}{2}|(0)}_{s} = \sqrt{[\hat{{\bPsi}}_{x_\textsl{e}}]_{1,1}/N}\,\mathbf{1}$, or based on the comparably lowly complex estimator in (\ref{eq:sota:phixeHatPrime_minprob})--(\ref{eq:sota:phixeHat}), here denoted by $\hat{\bphi}{}_{s|\textsl{c}_0}$, as $\hat{\bphi}{}^{\nicefrac{1}{2}|(0)}_{s} = \sqrt{\hat{\bphi}{}_{s|\textsl{c}_0}}$.
Here, the latter provides faster convergence, cf. Sec. \ref{sec:sim:modeled:results}.

\subsection{Recursive RETF Update}
\label{sec:RETFup}

Based upon the square-root model in (\ref{eq:fact:Psixe_fact}), the estimates $\hat{\bOmega}$ and $\hat{\bphi}{}^{\nicefrac{1}{2}}_{s}$ obtained as discussed in Sec. \ref{sec:oppPSD} may also be used to recursively update the RETF estimate $\hat{\H}$.
In the following, we differentiate the prior and posterior estimates $\hat{\H}$ and $\hat{\H}^{\scriptscriptstyle +}$, 
and propose to simply propagate the posterior in the previous frame to the prior in the current frame, i.e. 
\begin{align}
\hat{\H}(l) = \hat{\H}^{\scriptscriptstyle +}(l\-1). 
\label{eq:RETFup:Hprior}
\end{align}
In each frame, we use $\hat{\H}$ to obtain $\hat{\bOmega}$ and $\hat{\bphi}{}^{\nicefrac{1}{2}}_{s}$ with (\ref{eq:oppPSD:minprobOmega})--(\ref{eq:oppPSD:minprobPhi}), and then use  $\hat{\bOmega}$ and $\hat{\bphi}{}^{\nicefrac{1}{2}}_{s}$  to obtain $\hat{\H}^{\scriptscriptstyle +}$, where we again resort to the square-root signal model in (\ref{eq:fact:Psixe_fact}).
To this end, we define the approximation error as a function of $\H$,
\begin{align}
\mathbf{E}_{\textsl{sq}}(\H) &= \hat{{\bPsi}}{}^{\nicefrac{1}{2}}_{x_\textsl{e}}\hat{\bOmega} -  {\H}\Diag[ \hat{\bphi}{}^{\nicefrac{1}{2}}_{s}]
\label{eq:RETFup:Ef}
\end{align}
which is similar to (\ref{eq:oppPSD:Ef}). 
Based upon (\ref{eq:RETFup:Ef}) and the constraint in (\ref{eq:sm:H_constr}), 
we define the minimization problem,
 \begin{empheq}[box=\fbox]{align}
&\hat{\H}{}^{\scriptscriptstyle +} = \nonumber\\
&\quad \underset{\H}{\arg\min}\,
	\bigl\Vert\mathbf{E}_{\textsl{sq}}(\H)\bigr\Vert^2_\textsl{F}
	+ \bigl\Vert (\hat{\H} - \H)\Diag\bigl[\sqrt{\boldsymbol{\beta}}\,\bigr] \bigr\Vert_\textsl{F}^2 \nonumber\\
&\quad\hphantom{\arg}\st \quad\mathbf{i}^{\transp}\H = \mathbf{1}^{\transp},
\label{eq:RETFup:minprob}
\end{empheq}
%
where the penalty term $\bigl\Vert (\hat{\H} - \H)\Diag\bigl[\displaystyle{\sqrt{\boldsymbol{\beta}}}\,\bigr] \bigr\Vert_\textsl{F}^2$ relates to Leven-berg-Marquardt regularization \cite{vanWaterschootToon2008Oraf, LjungLennart1986Tapo} in that it penalizes deviation from the previous (i.e., the prior) estimate $\hat{\H}$. 
Here, we leave  $\boldsymbol{\beta}$ subject to tuning, as outlined below.
In this respect, recall that according to (\ref{eq:fact:Psixe_fact}), both ${{\bPsi}}{}^{\nicefrac{1}{2}}_{x_\textsl{e}}$ and ${\H}$ span the same column space. 
However, due to modeling and estimation errors, this is not necessarily true for the corresponding estimates $\hat{{\bPsi}}{}^{\nicefrac{1}{2}}_{x_\textsl{e}}$ and $\hat{\H}$.
In particular, if the $n^{\text{th}}$ source image has a comparably low early PSD ${\varphi}{}_{s_n}$ or is inactive, then the associated subspace dimension will not be well or not at all be represented in $\hat{{\bPsi}}{}^{\nicefrac{1}{2}}_{x_\textsl{e}}$, and 
both $[\hat{\bOmega}{}_{s}]_{:,n}$ and $\hat{\varphi}{}^{\nicefrac{1}{2}}_{s_n}$ 
may exhibit comparably large estimation errors.
Further, the estimate $\hat{\varphi}{}^{\nicefrac{1}{2}}_{s_n}$ may contain residual late reverberation due to erroneous separation of $\hat{\bPsi}_{x}$ into $\hat{\bPsi}_{x_\textsl{e}}$ and ${\bPsi}_{x_\ell}$, cf. (\ref{eq:sm:Psiy}), Sec. \ref{sec:GEVD} and Sec. \ref{sec:sim}.
In such a case, one would preferably rely on the prior estimate $\hat{\h}_n$ instead of updating based on $[\hat{\bOmega}{}_{s}]_{:,n}$ and $\hat{\varphi}{}^{\nicefrac{1}{2}}_{s_n}$.
Considering the solution to (\ref{eq:RETFup:minprob}), which is given by
 \begin{align} 
[\hat{\H}{}^{\scriptscriptstyle +}]_{1,:} &= \mathbf{1}^T,\\
[\hat{\H}{}^{\scriptscriptstyle +}]_{2:M,:} &=  \bigl{[(}\hat{{\bPsi}}{}^{\nicefrac{1}{2}}_{x_\textsl{e}}\hat{\bOmega}\Diag[ \hat{\bphi}{}^{\nicefrac{\herm}{2}}_{s}]  + \hat{\H}\Diag[\boldsymbol{\beta}]\bigr)\nonumber\\
&\qquad\cdot\Diag^{\inv}[\hat{\bphi}{}_{s} + \boldsymbol{\beta}]\bigr]_{2:M,:},
\label{eq:oppPSD:}
\end{align}
we indeed find that the smaller ${\hat{\varphi}}{}_{s_n}$ as compared to $\beta_n = [\boldsymbol{\beta}]_n$, the more $\hat{\h}{}^{\scriptscriptstyle +}_n$ relies on $\hat{\h}_n$, as desired. 
In order to further increase robustness against modeling and estimation errors, source inactivity and residual late reverberation in ${\hat{\varphi}}{}_{s_n}$, we propose to make $\beta_n$ time-varying with binary values.
More precisely, we base $\beta_n$ on the power ratio
\begin{align}
{\boldsymbol{\xi}} &= \hat{\bphi}{}_{s}/(\mathbf{1}^{\transp}\hat{\bphi}{}_{s} + \varphi_\textsl{reg}),
\label{eq:RETFup:xi}
\end{align}
where ${\xi}_n = [\boldsymbol{\xi}]_n \in [0,\,1]$. 
Here, $\varphi_\textsl{reg}$ may be used for regularization, 
e.g., we may choose $\varphi_\textsl{reg} = {\varphi}_{x_\ell}$ in order to limit ${{\xi}}_n$ in frames where pre-dominantly late reverberation is estimated.
Given ${\xi}_n$, we set ${\beta}_n$ as
\begin{align}
{\beta}_n & 
\begin{cases}
= \beta \quad &\text{if} \quad  {\xi}_n \geq \xi_\textsl{th},\\
\rightarrow \infty \quad &\text{else},
\end{cases}
\label{eq:RETFup:w}
\end{align}
and thereby resort to $\hat{\h}{}^{\scriptscriptstyle +}_n = \hat{\h}_n$ if $\xi_n$ is smaller than the pre-defined threshold $\xi_\textsl{th}$.
The value $\beta$, used if $\xi_n \geq \xi_\textsl{th}$, should scale in relation to the dynamic range of $\varphi_{s_n}$ and may be chosen depending on the (estimated) probability density function of the complex STFT coefficients $s_n$, cf. Sec \ref{sec:sim}.

Note that in order to start the recursion defined by (\ref{eq:RETFup:Hprior}), (\ref{eq:oppPSD:minprobOmega})--(\ref{eq:oppPSD:minprobPhi}), and (\ref{eq:RETFup:minprob}), an initial estimate $\hat{\H}(0)$ is required, which may be based on, e.g., initial single-source RETF estimates acquired from segments with mutual-exclusively active sources \cite{markovichAug09}, or some initial knowledge or estimates of the associated DoAs \cite{Thiergart2013AiMf, scheuing2008correlation, chen2010introduction}.

\section{Subspace-based Early Correlation Matrix Estimation}
\label{sec:GEVD}
In Sec. \ref{sec:sota} and Sec. \ref{sec:proposed}, we respectively assumed that the early-correlation-matrix estimate $\hat{\bPsi}_{x_\textsl{e}}$ and its square root $\hat{\bPsi}{}^{\nicefrac{1}{2}}_{x_\textsl{e}}$  of rank $N$ are available.
In this section, we discuss how to obtain these estimates from the microphone signals $\x$.
We estimate ${\bPsi}_{x}  = \Exp[\x\x^\herm]$ by recursively averaging $\x\x^\herm$, yielding the \textit{smooth} estimate $\hat{\bPsi}_{x|\textsl{sm}}$ and its equally \textit{smooth} subspace representation based on the GEVD.
From the latter, we first define a \textit{desmoothed} estimate $\hat{\bPsi}_{x}$, and second extract the early component $\hat{\bPsi}_{x_\textsl{e}}$ and its square root $\hat{\bPsi}{}^{\nicefrac{1}{2}}_{x_\textsl{e}}$.

In Sec. \ref{sec:smoothGEVD}, we introduce the subspace model of ${\bPsi}_{x}$.
In Sec. \ref{sec:desmoothGEVD}, we obtain the smooth and desmoothed estimates $\hat{\bPsi}_{x|\textsl{sm}}$ and $\hat{\bPsi}_{x}$, respectively.
In Sec. \ref{sec:decompGEVD}, given $\hat{\bPsi}_{x}$, we then retrieve a subspace-based rank-$N$ estimates $\hat{\bPsi}_{x_\textsl{e}}$ and $\hat{\bPsi}{}^{\nicefrac{1}{2}}_{x_\textsl{e}}$.

\subsection{Correlation Matrix Subspace Decomposition}
\label{sec:smoothGEVD}

In each frame $l$, we define the GEVD \cite{Serizel2014} of ${\bPsi}_{x}$ and the diffuse coherence matrix $\mathbf{\Gamma}$, cf. (\ref{eq:sm:Psixl}), i.e.
\begin{align}
{\bPsi}_{x}\P &= \mathbf{\Gamma}\P{\mathbf{\Lambda}}_{x},
\label{eq:smoothGEVD:PsiyHatGEVD}\\
\text{with}\quad {\mathbf{\Lambda}}_{x} &= \Diag[{\boldsymbol{\lambda}}_{x}],
\label{eq:smoothGEVD:Lambday}
\end{align}
where ${\boldsymbol{\lambda}}_{x} \in \mathbb{R}^{M}$ comprises the generalized eigenvalues, and the columns of $\P \in \mathbb{C}^{M\times M}$ comprise the associated generalized eigenvectors. 
In the GEVD, the generalized eigenvectors in $\P$ are uniquely defined up to a scaling factor, and for any factorization $\mathbf{\Gamma} = \mathbf{\Gamma}^{\nicefrac{1}{2}}\mathbf{\Gamma}^{\nicefrac{\herm}{2}}$, we find that $\mathbf{\Gamma}^{\nicefrac{\herm}{2}}\P$ is {column-wise orthogonal} due to $\hat{\bPsi}_{x|\textsl{sm}}$ being Hermitian. 
In the following, without loss of generality, we assume the eigenvectors to be scaled such that $\mathbf{\Gamma}^{\nicefrac{\herm}{2}}\P$ is {unitary}, i.e.
\begin{align}
\P^\herm\mathbf{\Gamma}\P &= \I,
\label{eq:smoothGEVD:P_scaling}
\end{align}
and therefore, combining (\ref{eq:smoothGEVD:PsiyHatGEVD}) and (\ref{eq:smoothGEVD:P_scaling}),
\begin{align}
 \P^\herm{\bPsi}_{x}\P = {\mathbf{\Lambda}}_{x}.  \label{eq:smoothGEVD:eigvalues}
 \end{align}
An alternative, but mathematically equivalent formulation to the GEVD in (\ref{eq:smoothGEVD:PsiyHatGEVD}) is given by the EVD of the pre-whitened matrix ${{\bPsi}}{}'_{x} = \mathbf{\Gamma}^{-\nicefrac{1}{2}}{\bPsi}_{x} \mathbf{\Gamma}^{-\nicefrac{\herm}{2}}$ \cite{markovichAug09, MarkovichGolanShmulik2015Paot, KodrasiD18}, which is defined by ${{\bPsi}}{}'_{x}{\P}{}' = {\P}{}'{\mathbf{\Lambda}}'_{x}$.
By comparison with (\ref{eq:smoothGEVD:PsiyHatGEVD}), we find ${\mathbf{\Lambda}}'_{x} = {\mathbf{\Lambda}}_{x}$ and ${\P}' = \mathbf{\Gamma}^{\nicefrac{\herm}{2}}\P$, provided that the respective (generalized) eigenvalues are sorted in the same order, and the (generalized) eigenvectors scaled accordingly.

For convenience of presentation, assume that the generalized eigenvalues in ${\boldsymbol{\lambda}}_x$ are sorted in a descending order, and the generalized eigenvectors in $\P$ are sorted accordingly.
Then, inserting ${\bPsi}_{x} = {\bPsi}_{x_\textsl{e}} + {\bPsi}_{x_\ell}$ with ${\bPsi}_{x_\ell} = {\varphi}_{x_\ell}\mathbf{\Gamma}$, cf. (\ref{eq:sm:Psiy}) and (\ref{eq:sm:Psixl}), into (\ref{eq:smoothGEVD:eigvalues}) while making use of (\ref{eq:smoothGEVD:P_scaling}) yields
\begin{align}
{\mathbf{\Lambda}}_x &= \P^\herm{\bPsi}_{x_\textsl{e}}\P +  {\varphi}_{x_\ell}\I, \label{eq:decompGEVD:LambdayHat}
\end{align}
wherein ${\bPsi}_{x_\textsl{e}}$ and in consequence $\P^\herm{\bPsi}_{x_\textsl{e}}\P$ generally have rank $N$, and the latter in addition is diagonal, i.e. if $N<M$ we have
\begin{align}
\P^\herm{\bPsi}_{x_\textsl{e}}\P &=
\begin{pmatrix} 
{{\bLambda}}_{x_\textsl{e}} & \0\\
\0 & \0\\
\end{pmatrix}, \label{eq:decompGEVD:PPsiP}\\
\text{with}\quad {{\bLambda}}_{x_\textsl{e}} &= \Diag[\boldsymbol{\lambda}_{x_\textsl{e}}],\label{eq:decompGEVD:LambdaxeHat}
\end{align}
and $\boldsymbol{\lambda}_{x_\textsl{e}} \in \mathbb{R}^{N}$.

\subsection{Recursive Correlation Matrix Estimation and Desmoothing}
\label{sec:desmoothGEVD}

We compute a smooth estimate $\hat{\bPsi}_{x|\textsl{sm}}$ of ${\bPsi}_{x}$ by recursively averaging $\x\x^\herm$ using some pre-defined forgetting factor $\zeta \in (0,\,1)$, i.e.
\begin{align}
\hat{\bPsi}_{x|\textsl{sm}}(l) = \zeta \hat{\bPsi}_{x| \textsl{sm}}(l\-1) + (1\-\zeta)\x(l)\x^\herm(l),\label{eq:smoothGEVD:PsiyHat}
\end{align}
and perform the GEVD $\hat{\bPsi}_{x|\textsl{sm}}\hat{\P} = \mathbf{\Gamma}\hat{\P}\hat{\mathbf{\Lambda}}_{x|\textsl{sm}}$ similar to (\ref{eq:smoothGEVD:PsiyHatGEVD})--(\ref{eq:smoothGEVD:eigvalues}), 
with $\hat{\P}$ an estimate of $\P$ and $\hat{\mathbf{\Lambda}}_{x|\textsl{sm}} = \Diag[\hat{\boldsymbol{\lambda}}_{x|\textsl{sm}}]$ a smooth estimate of  ${\mathbf{\Lambda}}_{x}$. 
Note that in order to excite all subspace dimensions and the associated generalized eigenvalues and hence to achieve a meaningful decomposition,  $\hat{\bPsi}_{x|\textsl{sm}}$ needs to be well-conditioned, and so  $\zeta$ must be sufficiently close to one.
As discussed in Sec. \ref{sec:sm}, the PSDs $\varphi_{s_{n}}$ and ${\varphi}_{x_\ell}$ may be highly non-stationary,
while the associated coherence matrices $\h_n\h_n^\herm$ and  $\mathbf{\Gamma}$ are commonly assumed to be comparably slowly time-varying or even time-invariant.
In theory, a linear combination of the PSDs $\varphi_{s_{n}} $ and ${\varphi}_{x_\ell}$ is rendered by the \textit{unknown} generalized eigenvalues  $\boldsymbol{\lambda}_{x}$ of ${\bPsi}_{x}$ and $\mathbf{\Gamma}$, i.e. also $\boldsymbol{\lambda}_{x}$ may be highly non-stationary.
In contrast, due to the (inevitable) recursive averaging in (\ref{eq:smoothGEVD:PsiyHat}), the \textit{computed} generalized eigenvalues $\hat{\boldsymbol{\lambda}}_{x|\textsl{sm}}$ of $\hat{\bPsi}_{x|\textsl{sm}}$ and $\mathbf{\Gamma}$
are slowly time-varying if  $\zeta$  is sufficiently large, i.e. non-stationarities are to some extent smoothed, and so would be PSD estimates
based on $\hat{\boldsymbol{\lambda}}_{x|\textsl{sm}}$ or $\hat{\bPsi}_{x|\textsl{sm}}$.
While smooth PSD estimates are commonly preferred in some applications (e.g., for perceptual reasons, in the computation of spectral gains in speech enhancement \cite{loizou2007speech}), others may exploit non-stationarities (such as, e.g., the Kalman filter \cite{haykin02}, where PSD estimates of the observation noise act as a regularization term in the recursive update of the state estimate \cite{dietzen19TASLP_ISCLP}).
Depending on the application, we therefore propose to restore non-stationarities by desmoothing $\hat{\boldsymbol{\lambda}}_{x|\textsl{sm}}$, yielding an estimate $\hat{\boldsymbol{\lambda}}_{x}$ of $\boldsymbol{\lambda}_{x}$.

To this end, we note that the recursive averaging in (\ref{eq:smoothGEVD:PsiyHat}) may be considered an element-wise filtering operation with $\x(l)\x^\herm(l)$ as the input, $\hat{\bPsi}_{x|\textsl{sm}}(l)$ as the output, and the (all-pole) ${z}$-domain transfer function given by $(1-\zeta)/(1- \zeta z^{\inv})$.
Therefore, in order to desmooth $\hat{\boldsymbol{\lambda}}_{x|\textsl{sm}}(l)$, we propose to apply the corresponding (all-zero) inverse transfer function  given by  $(1- \zeta z^{\inv})/(1-\zeta)$ followed by non-negative thresholding, i.e.
\begin{align}
\hat{\boldsymbol{\lambda}}{}'_{x}(l) &= \frac{\hat{\boldsymbol{\lambda}}_{x|\textsl{sm}}(l) - \zeta \hat{\boldsymbol{\lambda}}_{x|\textsl{sm}}(l\-1)}{1-\zeta}, \label{eq:desmoothGEVD:lambdayHatPrime_desmooth}\\
\hat{\boldsymbol{\lambda}}_{x}(l) &= \text{max}[\hat{\boldsymbol{\lambda}}{}'_{x}(l),\, 
\mathbf{0}
], \label{eq:desmoothGEVD:lambdayHatPrime_desmooth_thresh}
\end{align}
where the thresholding in (\ref{eq:desmoothGEVD:lambdayHatPrime_desmooth_thresh}) avoids negative eigenvalue estimates, which otherwise may appear in a limited number of frames due to modeling and estimation errors.
Note that the desmoothing operation requires the associated generalized eigenvalues in $\hat{\boldsymbol{\lambda}}_{x|\textsl{sm}}(l)$ and $\hat{\boldsymbol{\lambda}}_{x|\textsl{sm}}(l\-1)$ to be sorted correspondingly.
This may be ensured by sorting $\hat{\P}(l)$ such that $\hat{\P}^\herm(l\-1)\mathbf{\Gamma}\hat{\P}(l)\approx \I$, cf. (\ref{eq:smoothGEVD:P_scaling}), and $\hat{\boldsymbol{\lambda}}_{x|\textsl{sm}}(l)$ accordingly, which can be done easily for large $\zeta$ and the therewith slowly time-varying GEVD \cite{taslp19aCode}. 
Alternatively, recursive sorting may be avoided if the GEVD is estimated recursively, e.g., by means of the power method \cite{golub2012matrix, TammenMarvin2018CRoE}.
One may then define the corresponding desmoothed estimate $\hat{\bPsi}_{x}$ via its decomposition
\begin{align}
\hat{\bPsi}_{x}\hat{\P} &= \mathbf{\Gamma}\hat{\P}\hat{\mathbf{\Lambda}}_{x},\label{eq:desmoothGEVD:PsiyHatGEVD_desmooth}\\
\text{with}\quad \hat{\mathbf{\Lambda}}_{x} &= \Diag[\hat{\boldsymbol{\lambda}}_{x}],
\label{eq:desmoothGEVD:LambdayHat_desmooth}
\end{align}
where $\hat{\P}$ remains unchanged.

\subsection{Early Correlation Matrix Estimation and Factorization}
\label{sec:decompGEVD}

Given  $\hat{\P}$ and $\hat{\mathbf{\Lambda}}_{x}$ in (\ref{eq:desmoothGEVD:PsiyHatGEVD_desmooth})--(\ref{eq:desmoothGEVD:LambdayHat_desmooth}), we now a retrieve the subspace-based rank-$N$ estimates $\hat{\bPsi}_{x_\textsl{e}}$ and $\hat{{\bPsi}}{}^{\nicefrac{1}{2}}_{x_\textsl{e}}$.
To this end, based on (\ref{eq:decompGEVD:LambdayHat})--(\ref{eq:decompGEVD:LambdaxeHat}), we note that $\boldsymbol{\lambda}_{x_\textsl{e}}$ may be estimated as
\begin{align}
\hat{\boldsymbol{\lambda}}_{x_\textsl{e}} &=  [\hat{\boldsymbol{\lambda}}_{x}]_{1:N} - \hat{\varphi}_{x_\ell}\mathbf{1}, \label{eq:decompGEVD:lambdaxeHat}
\end{align}
where $\hat{\varphi}_{x_\ell}$ in turn may be obtained by averaging the last $M-N$ generalized eigenvalues in $[\hat{\boldsymbol{\lambda}}_{x}]_{N+1:M}$ \cite{KodrasiD18}.
Considering (\ref{eq:decompGEVD:PPsiP})--(\ref{eq:decompGEVD:LambdaxeHat}), given ${\hat{\bLambda}}_{x_\textsl{e}} = \Diag[\hat{\boldsymbol{\lambda}}_{x_\textsl{e}}]$ from (\ref{eq:decompGEVD:lambdaxeHat}) and $\hat{\P}^{\inv} = \hat{\P}^{H} \mathbf{\Gamma}$ from (\ref{eq:smoothGEVD:P_scaling}), we can define a rank-$N$ estimate of ${\bPsi}_{x_\textsl{e}}$ as
\begin{align}
\hat{{\bPsi}}_{x_\textsl{e}} &= 
 \mathbf{\Gamma}\hat{\P}
 \begin{pmatrix} 
{\hat{\bLambda}}_{x_\textsl{e}} & \0\\
\0 & \0\\
\end{pmatrix}
 \hat{\P}^\herm\mathbf{\Gamma}\nonumber\\
&= \mathbf{\Gamma}[\hat{\P}]_{:,1:N}\hat{\bLambda}_{x_\textsl{e}}[\hat{\P}]_{:,1:N}^\herm\mathbf{\Gamma}. \label{eq:decompGEVD:PsixeHat}
\end{align}
From (\ref{eq:decompGEVD:PsixeHat}), we can further easily derive a square root $\hat{{\bPsi}}{}^{\nicefrac{1}{2}}_{x_\textsl{e}}$ as
\begin{align}
\hat{{\bPsi}}{}^{\nicefrac{1}{2}}_{x_\textsl{e}} &= \mathbf{\Gamma}[\hat{\P}]_{:,1:N}\hat{\bLambda}{}^{\nicefrac{1}{2}}_{x_\textsl{e}}
\label{eq:decompGEVD:PsixeHat_fact}\\
\text{with}\quad \hat{\bLambda}{}^{\nicefrac{1}{2}}_{x_\textsl{e}} &= \Diag[\hat{\boldsymbol{\lambda}}{}^{\nicefrac{1}{2}}_{x_\textsl{e}}],
 \label{eq:decompGEVD:LambdaxeHat_fact}
\end{align}
with arbitrary complex arguments of the elements in $\hat{\boldsymbol{\lambda}}{}^{\nicefrac{1}{2}}_{x_\textsl{e}}$. 

Note that as opposed to the order presented in Sec. \ref{sec:desmoothGEVD} and this section, we may also apply desmoothing only after obtaining a smooth estimate of the early correlation matrix and its square root, which showed to yield comparable results in our simulations.

\section{Simulations}
\label{sec:sim}

In this section, we compare the algorithms based on the conventional and the square-root MP as presented in Sec. \ref{sec:sota} and Sec. \ref{sec:proposed}, respectively.
We assume that an (initial) RETF estimate $\hat{\H}$ is available, and that $\hat{\bPsi}_{x_\textsl{e}}$ and $\hat{\bPsi}{}^{\nicefrac{1}{2}}_{x_\textsl{e}}$ are obtained as described in Sec. \ref{sec:GEVD}.

Apart from estimation errors in $\hat{\bPsi}_{x_\textsl{e}}$, $\hat{\bPsi}{}^{\nicefrac{1}{2}}_{x_\textsl{e}}$, and $\hat{\H}$, the performance of both algorithms is subject to modeling errors, cf. Sec. \ref{sec:sm}.
Unfortunately, due to the model deficiencies in (\ref{eq:sm:Psiy})--(\ref{eq:sm:phid}), exact and observable ground truth early PSDs ${{\bphi}{}_{s}}$ and ground truth RETFs ${\H}$ do not exist in a practical setup based on realistic acoustic data.
Therefore, in order to yield a broader understanding of the algorithms' behavior, we perform two kinds of simulations. 
In the first kind, instead of generating time-domain data and estimating ${\bPsi}_x$ in the STFT domain, we generate $\hat{\bPsi}_x = {\bPsi_x}$ directly based on (\ref{eq:sm:Psiy})--(\ref{eq:sm:Psixl}) and assumed geometric and physical properties, i.e. $\hat{\bPsi}_x$ is free of modeling and estimation errors.
This way, we are able to define exact ground truth early PSDs ${{\bphi}{}_{s}}$ and ground truth RETFs ${\H}$ that may be used to define exact performance measures.
Further, the estimates $\hat{\bPsi}_{x_\textsl{e}}$ and $\hat{\bPsi}{}^{\nicefrac{1}{2}}_{x_\textsl{e}}$ obtained as described in Sec. \ref{sec:GEVD} will be free of estimation errors, such that the performance of both algorithms depends on the RETF estimation error in $\hat{\H}$ and the algorithmic settings in Sec. \ref{sec:sota} and Sec. \ref{sec:proposed} only.
We refer to these simulations as the \textit{model-based-data} case.
In the second kind of simulations, we generate acoustic data in the time domain from recorded speech signals and measured room impulse responses (RIRs), and estimate ${\bPsi}_x$ in the STFT domain.
This way, the setup becomes more practical, however, evaluation becomes less trivial in terms of the definition of performance measures,  such that we need to define and rely on an approximate ground truth early PSD ${\tilde{\bphi}{}_{s}}$ as a reference. 
We refer to these simulations as the \textit{acoustic-data} case.
The model-based-data case and the acoustic-data case are discussed in Sec. \ref{sec:sim:modeled} and Sec. \ref{sec:sim:acoustic}, respectively.

\subsection{Model-based Data}
\label{sec:sim:modeled}

We define our performance measures in Sec. \ref{sec:sim:modeled:measures}, discuss the data-generation in Sec. \ref{sec:sim:modeled:datagen}, the algorithmic settings in Sec. \ref{sec:sim:modeled:algo}, and the evaluation results in Sec. \ref{sec:sim:modeled:results}.

\subsubsection{Performance Measures}
\label{sec:sim:modeled:measures}
We define the RETF estimation error,
\begin{align}
 {\mathbf{E}}_{H} &= \hat{\mathbf{H}} - \mathbf{H}, \label{eq:sim:EH}
 \end{align}
where $\mathbf{i}^{\transp} {\mathbf{E}}_{H} = [ {\mathbf{E}}_{H}]_{1,:} = \mathbf{0}^{\transp}$ since both ${\mathbf{H}}$ and $\hat{\mathbf{H}}$ satisfy (\ref{eq:sm:H_constr}), 
and based on that the relative squared RETF estimation error,
\begin{align}
\varepsilon_{{H}} &= 10\log_{10}\frac{\tr[ {\mathbf{E}}_{H}^\herm{\mathbf{E}}_{H}]}{\tr[\mathbf{H}^\herm\mathbf{H}]-N} \si{dB}, \label{eq:sim:epsH}
\end{align}
where we subtract $N$ in the denominator in order to compensate for the fact that the first row of  $\mathbf{H}$ is known.
Since the early PSDs $\bphi_{s}$ are already a second-order property of the underlying signal $\mathbf{s}$, we define the PSD estimation error with respect to the non-negative square root of $\hat{\bphi}_{s}$ and $\bphi_{s}$, i.e. 
\begin{align}
%
%
\mathbf{e}_{\varphi_s} &= \sqrt{\hat{\bphi}_{s}} - \sqrt{\vphantom{\hat{\bphi}_{s}}\bphi_{s}}, \label{eq:sim:ephis}
\end{align}
and based on that the relative squared PSD estimation error, 
\begin{align}
\varepsilon_{\varphi_s} &= 10\log_{10} \frac{\mathbf{e}^\transp_{\varphi_s}\mathbf{e}_{\varphi_s}}{\mathbf{1}^\transp{\bphi}{}_{s}}\, \si{dB}. \label{eq:sim:epsphis}
\end{align}

 \setlength\fwidth{7.95cm}
 \begin{figure}[t]
\centering
\hspace*{-0.21cm}
    \setlength\fheight{4cm} 
    \input{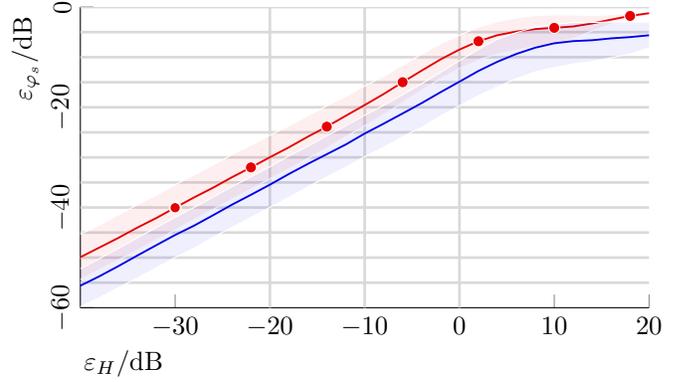} 
\caption{$\varepsilon_{{\varphi}{}_{s}}$ versus $\varepsilon_{{H}}$ for conventional MP [\ref{TASLP_OrthProc_overSNR_N3_laplacian_5}] and square-root  MP [\ref{TASLP_OrthProc_overSNR_N3_laplacian_8}] with $\alpha = 10^3$ at $f = 2\,$\si{kHz}.}
\label{fig:overepsH}
\end{figure}
 
  \setlength\fwidth{7.9cm}
 \begin{figure}[t]
\centering
\hspace*{-0.21cm}
    \setlength\fheight{1.64cm} 
    \input{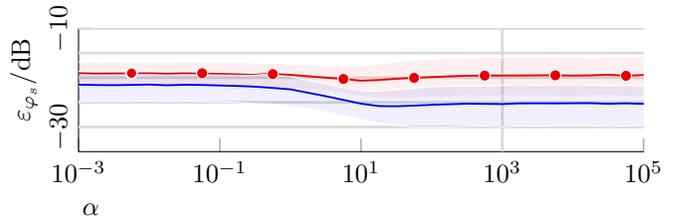} 
\caption{$\varepsilon_{{\varphi}{}_{s}}$ versus $\alpha$ for conventional MP [\ref{TASLP_OrthProc_overSNR_N3_laplacian_5}] and square-root MP [\ref{TASLP_OrthProc_overSNR_N3_laplacian_8}] at $\varepsilon_{{H}} = -10\,$\si{dB} and $f = 2\,$\si{kHz}.}
\label{fig:overalpha}
\end{figure}

\subsubsection{Data Generation}
\label{sec:sim:modeled:datagen}

Let $\hat{\bPsi}_x$ be available and free of modeling and estimation errors, i.e. we have $\hat{\bPsi}_x = {\bPsi_x}$ with $ {\bPsi_x}$ adhering to (\ref{eq:sm:Psiy})--(\ref{eq:sm:Psixl}).
We generate ${\bPsi}_x$ based on assumed geometric and physical properties.
We assume a linear microphone array of $M=5$ microphones with inter-microphone distance of 8$\,$\si{cm} and the speed of sound to be $340\,$\si{m/s}.
Further, we assume $N = 3$ sources, positioned at $(-30, 0, 60)^\circ$ relative to the broadside direction of the microphone array.
The RETFs $\H$ are generated assuming omnidirectional microphones of equal gain as well as free- and far-field propagation for the early components, i.e. $\H$ depends on the DoAs only and is fully defined by the corresponding phase shifts between microphones. 
The estimate $\hat{\H}$ is generated by adding an error component $\mathbf{E}_H$ according to ({\ref{eq:sim:EH}), where the elements $[\mathbf{E}_H]_{:,2:M}$ are drawn from independent complex Gaussian distributions,  yielding a particular $\varepsilon_H$ according to (\ref{eq:sim:epsH}).
The diffuse coherence matrix $\bGamma$ is computed assuming a spherical-isotropic sound field. 
The early PSDs ${{\bphi}{}_{s}}$ are generated in the following manner. 
We draw the real and imaginary parts of the elements of $\mathbf{s}$ from independent Laplace distributions, 
which is a commonly assumed distribution for STFT coefficients of speech \cite{Martin05, LotterV05}, 
i.e. we have $\Re [s_n] \sim (\nicefrac{1}{b})e^{{-2\vert\Re [s_n] \vert}/{b}}$ and $\Im [s_n] \sim (\nicefrac{1}{b})e^{{-2\vert\Im [s_n] \vert}/{b}}$, where the scaling parameter $b$ is referred to as diversity.
Then, we define ${{\bphi}{}_{s}} = \Diag[{\mathbf{s}}^\herm]{\mathbf{s}}$, i.e. ${{\bphi}{}_{s}}$ is the squared magnitude of ${\mathbf{s}}$.
Given the above, we set ${\bPsi}_{x_\textsl{e}} = \H{{\bphi}{}_{s}}\H^\herm$ according to (\ref{eq:sm:Psixe}).
Note that since $\hat{\bPsi}_x = {\bPsi}_x$ is free of modeling errors, where 
 ${\bPsi}_x = {\bPsi}_{x_\textsl{e}} + {\bPsi}_{x_\ell}$ with ${\bPsi}_{x_\ell} = {\varphi}_{x_\ell}\bGamma$, cf. (\ref{eq:sm:Psiy}) and (\ref{eq:sm:Psixl}), the component ${\bPsi}_{x_\textsl{e}}$ may be perfectly estimated from $\hat{\bPsi}_x$ by means of the GEVD as described in Sec. \ref{sec:decompGEVD}, yielding $\hat{\bPsi}_{x_\textsl{e}} = {\bPsi}_{x_\textsl{e}}$ independently of ${\varphi}_{x_\ell}$.
Further, note that next to $\H$ and ${{\bphi}{}_{s}}$, via the GEVD, also $\bGamma$ influences the shape of the square root $\hat{\bPsi}_{x_\textsl{e}}^{\nicefrac{1}{2}} = {\bPsi}_{x_\textsl{e}}^{\nicefrac{1}{2}}$ in the sense of defining the basis for a given vector space, cf. Sec. \ref{sec:decompGEVD}.
For each data-point in the evaluation, cf. Sec. \ref{sec:sim:modeled:results}, we simulate $2^{14}$ realizations of $\hat{\bPsi}_{x_\textsl{e}}$, $\hat{\bPsi}_{x_\textsl{e}}^{\nicefrac{1}{2}}$ and $\hat{\H}$. 

\subsubsection{Algorithmic Settings}
\label{sec:sim:modeled:algo}

In the model-based-data case, as opposed to the acoustic-data case, cf. Sec. \ref{sec:sim:acoustic:algorithmic}, the sampling frequency and STFT-processing parameters are irrelevant since we generate $\hat{\bPsi}_x$ directly in the STFT-domain, cf. Sec. \ref{sec:sim:modeled:datagen}.
Regardless, we simulate frequencies up to $f = 8\,$\si{kHz}, corresponding to a virtual sampling frequency of $f_s = 16\,$\si{kHz}.
The soft-constraint penalty factor $\alpha$ in the conventional MP in (\ref{eq:sota:phixeHat_minprob}) and the square-root MP in (\ref{eq:oppPSD:minprob}) is simulated in the range $\alpha \in [10^{-3}, 10^{5}]$.
We perform at most $i_\textsl{max} = 20$ iterations of the associated iterative algorithms in (\ref{eq:sota:proxgradgrad})--(\ref{eq:sota:proxgradthres}) and (\ref{eq:oppPSD:minprobOmega})--(\ref{eq:oppPSD:minprobPhi}).
All but one of our simulations consider a single frame $l$ only.
In the one simulation considering recursive behavior, we do not update  $\hat{\H}$ for the conventional MP, but we do update $\hat{\H}$ recursively for the square-root MP as described in Sec. \ref{sec:RETFup}. 
In the latter case, in (\ref{eq:RETFup:xi}), since $\hat{\bPsi}{}^{\nicefrac{1}{2}}_{x_\textsl{e}} = {\bPsi}{}^{\nicefrac{1}{2}}_{x_\textsl{e}}$ is free of modeling and estimation errors and therefore free of residual late reverberation, cf. Sec. \ref{sec:sim:modeled:datagen}, we set $\varphi_{\textsl{reg}} = 0$.
 In (\ref{eq:RETFup:w}), the threshold $\xi_\textsl{th}$ is set as $10\log_{10} \xi_\textsl{th} = -2\,\si{dB}$ and $\beta$ is set as $\beta = 20 b^2$, with $b$ the diversity of the Laplace distributions used in the generation of $\bphi_s$, cf. Sec. \ref{sec:sim:modeled:datagen}.

\subsubsection{Results}
\label{sec:sim:modeled:results}

\setlength\fwidth{7.65cm}
\begin{figure}[t]
\centering
\hspace*{-0.245cm}
    \setlength\fheight{7.26cm} 
    \input{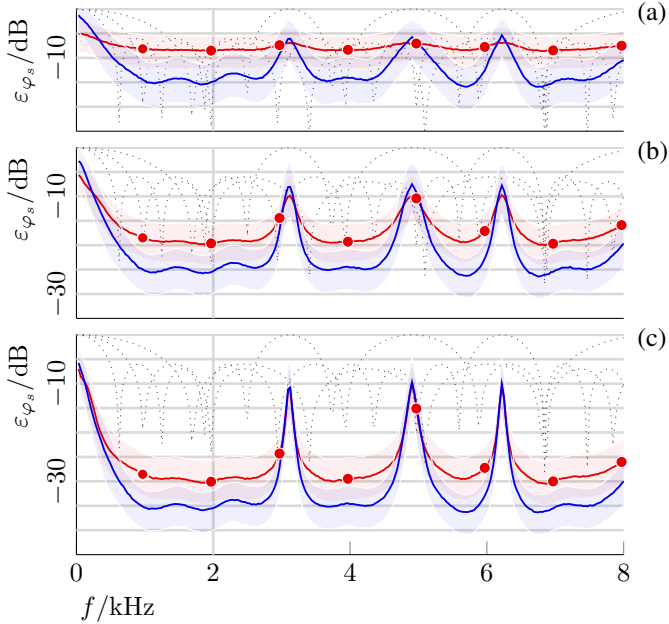} 
\caption{$\varepsilon_{{\varphi}{}_{s}}$ versus $f$ for  conventional MP [\ref{TASLP_OrthProc_overSNR_N3_laplacian_5}] and square-root MP [\ref{TASLP_OrthProc_overSNR_N3_laplacian_8}] with $\alpha = 10^3$ at (a) $\varepsilon_{{H}} = 0\,$\si{dB}, (b) $\varepsilon_{{H}} = -10\,$\si{dB}, and (c) $\varepsilon_{{H}} = -20\,$\si{dB}.
The graphs denoted by [\ref{TASLP_OrthProc_overFreq_N3_laplacian_1}] correspond to $10\,\text{log}_{10}\,{\lvert\h_n^\herm\h_{n'}\rvert}/{M} \,\text{\si{dB}}$ for $n'\neq n$.
}
\label{fig:overfreq}
\end{figure}

Fig. \ref{fig:overepsH} shows the PSD estimation performance in terms of the relative squared PSD estimation error $\varepsilon_{{\varphi}{}_{s}}$  for different values of the relative squared RETF estimation error $\varepsilon_{{H}}$ for the algorithms based on the conventional MP [\ref{TASLP_OrthProc_overSNR_N3_laplacian_5}] and the square-root MP [\ref{TASLP_OrthProc_overSNR_N3_laplacian_8}] with $\alpha = 10^3$ at $f = 2\,$\si{kHz} within a single frame $l$.
In this figure and similar ones in the following, the graphs denote medians over all  $2^{14}$ realizations, cf. Sec. \ref{sec:sim:modeled:datagen}, and the shaded areas denote the range from the first to the third quartile.
As can be seen, for both the conventional MP and the square-root MP, $\varepsilon_{{\varphi}{}_{s}}$ increases at a rate of about $10\,\si{dB}$ per $10\,\si{dB}$ increase in $\varepsilon_{{H}}$ until roughly  $\varepsilon_{{H}} = 0\,\si{dB}$ and $\varepsilon_{{H}} = 5\,\si{dB}$ is reached, respectively, after which $\varepsilon_{{\varphi}{}_{s}}$ begins to saturate.
This saturation is due to the fact that both algorithms yield non-negative estimates $\hat{\bphi}_s \geq \mathbf{0}$, which limits the estimation error at high values of $\varepsilon_{{H}}$.
The square-root MP outperforms the conventional MP by at least $5.7\,\si{dB}$ for $\varepsilon_{{H}} \leq 0\,\si{dB}$, and by somewhat less for $\varepsilon_{{H}} \geq 5\,\si{dB}$.

Fig. \ref{fig:overalpha} illustrates $\varepsilon_{{\varphi}{}_{s}}$ for different values of the soft constraint penalty factor $\alpha$ for the conventional MP [\ref{TASLP_OrthProc_overSNR_N3_laplacian_5}]  and the square-root MP [\ref{TASLP_OrthProc_overSNR_N3_laplacian_8}] at $\varepsilon_{{H}} = -10\,$\si{dB} and $f = 2\,$\si{kHz} within a single frame $l$.
We note that while $\alpha$ hardly impacts the performance of the conventional MP, we generally reach larger improvements for higher values of $\alpha$ in the square-root MP.
Recall  that the soft constraint in the conventional MP is scalar-based, cf. (\ref{eq:sota:ec}), while the soft constraint in the square-root MP is vector-based, cf. (\ref{eq:oppPSD:ef}), and is therefore more informative. 
The square-root MP outperforms the conventional MP by $2.5\,\si{dB}$ at low values of $\alpha$, and by $5.7\,\si{dB}$ at high values of $\alpha$.
Interestingly, for both algorithms, despite $\hat{\bPsi}_{x_\textsl{e}}$ and $\hat{\bPsi}{}^{\nicefrac{1}{2}}_{x_\textsl{e}}$ being free of estimation errors, the minimum of $\varepsilon_{{\varphi}{}_{s}}$ does not occur at the highest values of $\alpha$, but at around $\alpha = 10^1$. 
As compared to higher values, the improvement  is however mild.

\setlength\fwidth{7.22cm}
\begin{figure}[t]
\centering
\hspace*{-0.24cm}
    \setlength\fheight{10cm} 
    \input{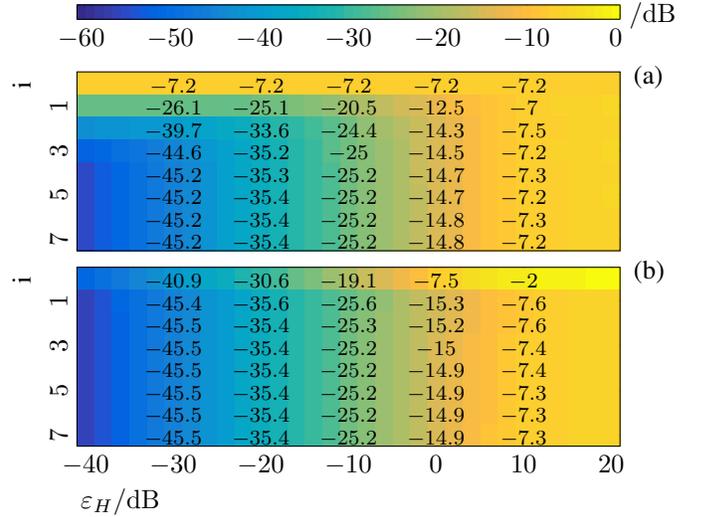} 
\caption{$\varepsilon^{(i)}_{{\varphi}{}_{s}}$ versus $\varepsilon_{{H}}$  and $i$ for square-root MP with $\alpha = 10^3$ and ${\hat{\bphi}{}^{\nicefrac{1}{2}|(0)}_{s}}$ based upon (a) the sum constraint in (\ref{eq:sm:phixe_constr}) and (b) the estimator in (\ref{eq:sota:phixeHatPrime_minprob})--(\ref{eq:sota:phixeHat}) at $f = 2\,$\si{kHz}.}
\label{fig:iterations}
\end{figure}

Fig. \ref{fig:overfreq} illustrates $\varepsilon_{{\varphi}{}_{s}}$ for different frequencies $f$ for the conventional MP [\ref{TASLP_OrthProc_overSNR_N3_laplacian_5}]  and the square-root MP [\ref{TASLP_OrthProc_overSNR_N3_laplacian_8}] with $\alpha = 10^3$ at (a) $\varepsilon_{{H}} = 0\,$\si{dB}, (b) $\varepsilon_{{H}} = -10\,$\si{dB}, and (c) $\varepsilon_{{H}} = -20\,$\si{dB} within a single frame $l$.
Note that at some frequencies, due to spatial aliasing, which occurs for two different DoAs if their phase difference in each microphone is a multiple of $2\pi$, the two corresponding DoA-based RETFs in $\H$, cf. Sec. \ref{sec:sim:modeled:datagen}, will be identical, and therefore $\H$ itself and consequently also $\bPsi_{x_\textsl{e}}$ and $\bPsi^{\nicefrac{1}{2}}_{x_\textsl{e}}$ will be rank-deficient. 
In our setup, this situation occurs for $f \in \{3.11, 4.91, 6.22\}\,\si{kHz}$, cf. also the dotted lines [\ref{TASLP_OrthProc_overFreq_N3_laplacian_1}] corresponding to $10\,\text{log}_{10}\,{\lvert\h_n^\herm\h_{n'}\rvert}/{M} \,\text{\si{dB}}$ for $n'\neq n$, which reach $0\,\si{dB}$ if $\h_{n'} = \h_n$.
As expected, by comparing Fig. \ref{fig:overfreq} (a) to Fig. \ref{fig:overfreq} (c), neither of the two algorithms performs well in the proximity of these frequencies, independent of $\varepsilon_{{H}}$.
At other frequencies, however, the square-root MP outperforms the conventional MP by roughly $5$ to $7\,\si{dB}$.

Fig. \ref{fig:iterations} demonstrates the effect in the median of the initial estimate $\hat{\bphi}{}_s^{\nicefrac{1}{2}|(0)}$ on the convergence behavior  in terms of the relative squared PSD estimation error $\varepsilon_{{\varphi}{}_{s}}^{(i)}$ at iteration $i$ for different values of $\varepsilon_{{H}}$ of the iterative algorithm in (\ref{eq:oppPSD:minprobOmega})--(\ref{eq:oppPSD:minprobPhi}) solving the square-root MP with $\alpha = 10^3$ at $f = 2\,$\si{kHz}.
The initial value is based on (a) the sum constraint in (\ref{eq:sm:phixe_constr}) as $\hat{\bphi}{}^{\nicefrac{1}{2}|(0)}_{s} = \sqrt{[\hat{{\bPsi}}_{x_\textsl{e}}]_{1,1}/N}\,\mathbf{1}$, and (b) the estimator in (\ref{eq:sota:phixeHatPrime_minprob})--(\ref{eq:sota:phixeHat}), here denoted by $\hat{\bphi}{}_{s|\textsl{c}_0}$, as $\hat{\bphi}{}^{\nicefrac{1}{2}|(0)}_{s} = \sqrt{\hat{\bphi}{}_{s|\textsl{c}_0}}$.
In both cases, the algorithm converges to almost the same final value of $\varepsilon_{{\varphi}{}_{s}}$.
However, we find that in (a), convergence is reached at around $i=3$ to $i=4$, while in (b), due to the improved initial estimate, convergence is reached at $i=1$ already.
Hence, while the computation of the initial estimate in (b) is somewhat more expensive, we save $2$ to $3$ iterations as compared to (a). 

Fig. \ref{fig:recursions} demonstrates the recursive behavior in terms of (a) $\varepsilon_{{H}}(l{\hskip 0.13em {+}\hskip 0.13em}r)$ and (b) $\varepsilon_{{\varphi}{}_{s}}(l{\hskip 0.13em {+}\hskip 0.13em}r)$ with $r$ the recursion index for the conventional MP [\ref{TASLP_OrthProc_overSNR_N3_laplacian_5}] and the square-root MP [\ref{TASLP_OrthProc_overSNR_N3_laplacian_8}] with $\alpha = 10^3$ at $f = 2\,$\si{kHz} and $\varepsilon_{{H}}(l) = 0\,\si{dB}$.  
Here, the source positioned at $-30^\circ$ transitions to $-40^\circ$ at $r = 32$, resulting in a transient change in the otherwise constant RETF $\H$.
While no update of the estimate $\hat{\H}$ is performed for the conventional MP, we do update $\hat{\H}$ recursively for  square-root MP as described in Sec. \ref{sec:RETFup}.
For the conventional MP, we expectably find that $\varepsilon_{{H}}(l{\hskip 0.13em {+}\hskip 0.13em}r)$ and $\varepsilon_{{\varphi}{}_{s}}(l{\hskip 0.13em {+}\hskip 0.13em}r)$ remain constant except for a transient increase of $6.8\,\si{dB}$ and $3.2\,\si{dB}$ at $r = 33$, respectively.
For the square-root MP, due to the recursive update of $\hat{\H}$, we find that $\varepsilon_{{H}}(l{\hskip 0.13em {+}\hskip 0.13em}r)$ and $\varepsilon_{{\varphi}{}_{s}}(l{\hskip 0.13em {+}\hskip 0.13em}r)$ decrease by $5.2\,\si{dB}$ and $4.7\si{dB}$ over the course of the first 32 recursions, followed by an increase of $11.2\,\si{dB}$ and $6.1\,\si{dB}$ at $r = 33$, respectively, and a subsequent decrease at roughly the same rate.

 \setlength\fwidth{7.65cm}
 \begin{figure}[t]
\centering
\hspace*{-0.26cm}
    \setlength\fheight{3.15cm} 
    \input{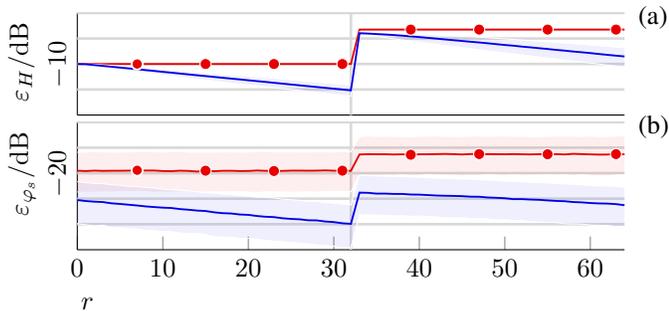} 
\caption{(a) $\varepsilon_{{H}}(l{\hskip 0.13em {+}\hskip 0.13em}r)$ and (b) $\varepsilon_{{\varphi}{}_{s}}(l{\hskip 0.13em {+}\hskip 0.13em}r)$ versus $r$ for
conventional MP [\ref{TASLP_OrthProc_overSNR_N3_laplacian_5}] and square-root MP [\ref{TASLP_OrthProc_overSNR_N3_laplacian_8}] with $\alpha = 10^3$ at $f = 2\,$\si{kHz} and $\varepsilon_{{H}}(l) = 0\,$\si{dB} if $\H$ changes at $r = 32$ and remains constant otherwise.}
\label{fig:recursions}
\end{figure}

\subsection{Acoustic Data}
\label{sec:sim:acoustic}

We define the performance measures in Sec. \ref{sec:sim:acoustic:measures}, discuss the acoustic scenario in Sec. \ref{sec:sim:acoustic:acoustscen}, the algorithmic settings in Sec. \ref{sec:sim:acoustic:algorithmic}, and the evaluation results in Sec. \ref{sec:sim:acoustic:results}.

 \setlength\fwidth{16cm}
 \begin{figure*}[t]
\centering
\hspace*{-0.205cm}
    \setlength\fheight{7.8cm} 
%
%
\begin{tikzpicture}

\begin{axis}[%
width=\fwidth,
height=0.022\fheight,
at={(0\fwidth,0.967\fheight)},
scale only axis,
point meta min=-55,
point meta max=5,
axis on top,
xmin=0.5,
xmax=61.5,
xtick={6,16,26,36,46,56},
xticklabels={{$-50$},{$-40$},{$-30$},{$-20$},{$-10$},{$0$}},
separate axis lines,
every outer y axis line/.append style={black},
every y tick label/.append style={font=\color{black}},
ymin=0.5,
ymax=2.5,
ytick={1},
yticklabels={{\rotatebox{-90}{$/$\si{dB}}}},
axis background/.style={fill=white},
yticklabel pos=right,
major tick length = 0em, ylabel style={at={(axis description cs:0.045,1)}, xshift=.0em, anchor=north east}, yticklabel style={rotate=90}, xlabel style={at={(ticklabel cs: -0.01,-8)}, anchor=west}, title style={at={(axis description cs:.92,0)}, yshift=-1.1em, anchor=south, font=\normalfont}
]
\addplot [forget plot] graphics [xmin=0.5,xmax=61.5,ymin=0.5,ymax=2.5] {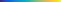};
\end{axis}

\begin{axis}[%
width=0.242\fwidth,
height=0.439\fheight,
at={(0\fwidth,0.45\fheight)},
scale only axis,
point meta min=-55,
point meta max=5,
axis on top,
xmin=0.5,
xmax=94.5,
xtick={\empty},
ymin=0.5,
ymax=257.5,
ytick={33,97,161,225},
yticklabels={{1},{3},{5},{7}},
ylabel={$f/\si{kHz}$},
axis background/.style={fill=white},
title style={font=\bfseries},
title={\colorbox{white}{(a.1)}},
major tick length = 0em, ylabel style={at={(axis description cs:0.045,1)}, xshift=.0em, anchor=north east}, yticklabel style={rotate=90}, xlabel style={at={(ticklabel cs: -0.01,-8)}, anchor=west}, title style={at={(axis description cs:.92,0)}, yshift=-1.1em, anchor=south, font=\normalfont}
]
\addplot [forget plot] graphics [xmin=0.5,xmax=94.5,ymin=0.5,ymax=257.5] {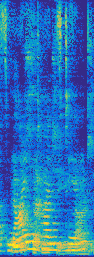};
\end{axis}

\begin{axis}[%
width=0.242\fwidth,
height=0.439\fheight,
at={(0\fwidth,0\fheight)},
scale only axis,
point meta min=-55,
point meta max=5,
axis on top,
xmin=0.5,
xmax=94.5,
xtick={1,64},
xticklabels={{0},{1}},
xlabel={$t/\si{s}$},
ymin=0.5,
ymax=257.5,
ytick={33,97,161,225},
yticklabels={{1},{3},{5},{7}},
ylabel={$f/\si{kHz}$},
axis background/.style={fill=white},
title style={font=\bfseries},
title={\colorbox{white}{(a.2)}},
major tick length = 0em, ylabel style={at={(axis description cs:0.045,1)}, xshift=.0em, anchor=north east}, yticklabel style={rotate=90}, xlabel style={at={(ticklabel cs: -0.01,-8)}, anchor=west}, title style={at={(axis description cs:.92,0)}, yshift=-1.1em, anchor=south, font=\normalfont}
]
\addplot [forget plot] graphics [xmin=0.5,xmax=94.5,ymin=0.5,ymax=257.5] {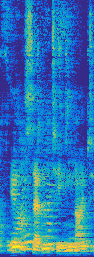};
\end{axis}

\begin{axis}[%
width=0.242\fwidth,
height=0.439\fheight,
at={(0.253\fwidth,0.45\fheight)},
scale only axis,
point meta min=-55,
point meta max=5,
axis on top,
xmin=0.5,
xmax=94.5,
xtick={\empty},
ymin=0.5,
ymax=257.5,
ytick={\empty},
axis background/.style={fill=white},
title style={font=\bfseries},
title={\colorbox{white}{(b.1)}},
major tick length = 0em, ylabel style={at={(axis description cs:0.045,1)}, xshift=.0em, anchor=north east}, yticklabel style={rotate=90}, xlabel style={at={(ticklabel cs: -0.01,-8)}, anchor=west}, title style={at={(axis description cs:.92,0)}, yshift=-1.1em, anchor=south, font=\normalfont}
]
\addplot [forget plot] graphics [xmin=0.5,xmax=94.5,ymin=0.5,ymax=257.5] {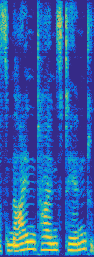};
\end{axis}

\begin{axis}[%
width=0.242\fwidth,
height=0.439\fheight,
at={(0.253\fwidth,0\fheight)},
scale only axis,
point meta min=-55,
point meta max=5,
axis on top,
xmin=0.5,
xmax=94.5,
xtick={1,64},
xticklabels={{0},{1}},
xlabel={$t/\si{s}$},
ymin=0.5,
ymax=257.5,
ytick={\empty},
axis background/.style={fill=white},
title style={font=\bfseries},
title={\colorbox{white}{(b.2)}},
major tick length = 0em, ylabel style={at={(axis description cs:0.045,1)}, xshift=.0em, anchor=north east}, yticklabel style={rotate=90}, xlabel style={at={(ticklabel cs: -0.01,-8)}, anchor=west}, title style={at={(axis description cs:.92,0)}, yshift=-1.1em, anchor=south, font=\normalfont}
]
\addplot [forget plot] graphics [xmin=0.5,xmax=94.5,ymin=0.5,ymax=257.5] {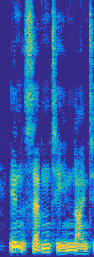};
\end{axis}

\begin{axis}[%
width=0.242\fwidth,
height=0.439\fheight,
at={(0.506\fwidth,0.45\fheight)},
scale only axis,
point meta min=-55,
point meta max=5,
axis on top,
xmin=0.5,
xmax=94.5,
xtick={\empty},
ymin=0.5,
ymax=257.5,
ytick={\empty},
axis background/.style={fill=white},
title style={font=\bfseries},
title={\colorbox{white}{(c.1)}},
major tick length = 0em, ylabel style={at={(axis description cs:0.045,1)}, xshift=.0em, anchor=north east}, yticklabel style={rotate=90}, xlabel style={at={(ticklabel cs: -0.01,-8)}, anchor=west}, title style={at={(axis description cs:.92,0)}, yshift=-1.1em, anchor=south, font=\normalfont}
]
\addplot [forget plot] graphics [xmin=0.5,xmax=94.5,ymin=0.5,ymax=257.5] {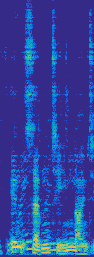};
\end{axis}

\begin{axis}[%
width=0.242\fwidth,
height=0.439\fheight,
at={(0.506\fwidth,0\fheight)},
scale only axis,
point meta min=-55,
point meta max=5,
axis on top,
xmin=0.5,
xmax=94.5,
xtick={1,64},
xticklabels={{0},{1}},
xlabel={$t/\si{s}$},
ymin=0.5,
ymax=257.5,
ytick={\empty},
axis background/.style={fill=white},
title style={font=\bfseries},
title={\colorbox{white}{(c.2)}},
major tick length = 0em, ylabel style={at={(axis description cs:0.045,1)}, xshift=.0em, anchor=north east}, yticklabel style={rotate=90}, xlabel style={at={(ticklabel cs: -0.01,-8)}, anchor=west}, title style={at={(axis description cs:.92,0)}, yshift=-1.1em, anchor=south, font=\normalfont}
]
\addplot [forget plot] graphics [xmin=0.5,xmax=94.5,ymin=0.5,ymax=257.5] {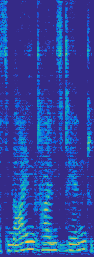};
\end{axis}

\begin{axis}[%
width=0.242\fwidth,
height=0.439\fheight,
at={(0.758\fwidth,0.45\fheight)},
scale only axis,
point meta min=-55,
point meta max=5,
axis on top,
xmin=0.5,
xmax=94.5,
xtick={\empty},
ymin=0.5,
ymax=257.5,
ytick={\empty},
axis background/.style={fill=white},
title style={font=\bfseries},
title={\colorbox{white}{(d.1)}},
major tick length = 0em, ylabel style={at={(axis description cs:0.045,1)}, xshift=.0em, anchor=north east}, yticklabel style={rotate=90}, xlabel style={at={(ticklabel cs: -0.01,-8)}, anchor=west}, title style={at={(axis description cs:.92,0)}, yshift=-1.1em, anchor=south, font=\normalfont}
]
\addplot [forget plot] graphics [xmin=0.5,xmax=94.5,ymin=0.5,ymax=257.5] {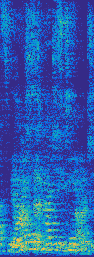};
\end{axis}

\begin{axis}[%
width=0.242\fwidth,
height=0.439\fheight,
at={(0.758\fwidth,0\fheight)},
scale only axis,
point meta min=-55,
point meta max=5,
axis on top,
xmin=0.5,
xmax=94.5,
xtick={1,64},
xticklabels={{0},{1}},
xlabel={$t/\si{s}$},
ymin=0.5,
ymax=257.5,
ytick={\empty},
axis background/.style={fill=white},
title style={font=\bfseries},
title={\colorbox{white}{(d.2)}},
major tick length = 0em, ylabel style={at={(axis description cs:0.045,1)}, xshift=.0em, anchor=north east}, yticklabel style={rotate=90}, xlabel style={at={(ticklabel cs: -0.01,-8)}, anchor=west}, title style={at={(axis description cs:.92,0)}, yshift=-1.1em, anchor=south, font=\normalfont}
]
\addplot [forget plot] graphics [xmin=0.5,xmax=94.5,ymin=0.5,ymax=257.5] {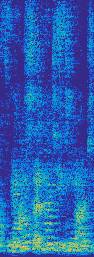};
\end{axis}
\end{tikzpicture}%
\caption{
Exemplary spectrograms depicting $\hat{\varphi}{}_{s_n}$ in (a.$n$),  $\bar{\varphi}{}_{s_n}$ in (b.$n$), ${e}_{\varphi_{s_n}}^{2|\textsl{int\,}}$ in (c.$n$), and ${e}_{\varphi_{s_n}}^{2|\textsl{art\,}}$ in (d.$n$),
with $\bar{\varphi}{}_{s_n}$, ${e}_{\varphi_{s_n}}^{2|\textsl{int\,}}$, and ${e}_{\varphi_{s_n}}^{2|\textsl{art\,}}$ obtained from the decomposition of $\sqrt{\hat{\varphi}{}_{s_n}\hspace{-2pt}}\hspace{2pt}$, cf. Sec. \ref{sec:sim:acoustic:measures}. 
The reference PSDs ${\tilde{\varphi}{}_{s_1}}$ and ${\tilde{\varphi}{}_{s_2}}$ 
originate from a female and a male speaker at $-30^\circ$ and $60^\circ$, respectively, and the estimate $\hat{\varphi}{}_{s_n}$ is obtained by means of the square-root MP.
}
\label{fig:spectrograms}
\end{figure*}

\subsubsection{Performance Measures}
\label{sec:sim:acoustic:measures}
In the acoustic-data case, due to the model deficiencies in (\ref{eq:sm:Psiy})--(\ref{eq:sm:phid}), cf. Sec. \ref{sec:sm}, exact and observable ground truth early PSDs ${{\bphi}{}_{s}}$ and ground truth RETFs ${\H}$ do unfortunately not exist, and so the performance measures in (\ref{eq:sim:EH})--(\ref{eq:sim:epsphis}) cannot be used.
However, one may define approximate ground truth early PSDs ${\tilde{\bphi}{}_{s}}$ as a reference for evaluation.
To this end, given the source signals and RIRs of a particular acoustic scenario, cf. Sec. \ref{sec:sim:acoustic:acoustscen}, we convolve the source signals with only the early part of the RIR to the first microphone and transform to the STFT-domain, yielding $\tilde{\mathbf{s}}$, and set ${\tilde{\bphi}{}_{s}} = \Diag[\tilde{\mathbf{s}}^\herm]\tilde{\mathbf{s}}$, i.e. ${\tilde{\bphi}{}_{s}}$ is the squared magnitude\footnote{
If  subspace-based desmoothing, cf. Sec. \ref{sec:desmoothGEVD}, is not applied in the computation of ${\hat{\bphi}{}_{s}}$, one may instead choose a recursively averaged version of the squared magnitude as a reference.
} of $\tilde{\mathbf{s}}$. 
Note that the definition of the early part of the RIR is somewhat arbitrary due to the weighted and overlapping windows in the STFT-processing.
For STFT windows of $N_\textsl{STFT}$ samples with $50\%$ overlap, 
one may, e.g., choose the first $N_\textsl{STFT}$ or the first $N_\textsl{STFT}/2$ taps of the RIR.
Here, we have chosen the first $N_\textsl{STFT}$ samples corresponding to $32\,\si{ms}$, cf. Sec. \ref{sec:sim:acoustic:algorithmic}.
In our setup, we have found that different choices result in quantitatively different performance, but not qualitatively different conclusions.

Given a segment of $L$ frames of ${\tilde{\bphi}{}_{s}}$ and ${\hat{\bphi}{}_{s}}$,
we decompose $\sqrt{\hat{\bphi}{}_{s}}$ according to \cite{VincentGF06} as
\begin{align}
\sqrt{\hat{\bphi}{}_{s}} =\sqrt{\vphantom{\hat{\bphi}_{s}}\bar{\bphi}_{s}} + \mathbf{e}_{\varphi_s} ^{\textsl{int\,}} + \mathbf{e}_{\varphi_s}^{\textsl{art\,}},  \label{eq:sim:decomp}
\end{align}
where $\sqrt{\vphantom{\hat{\varphi}{}_{s_n}}\bar{\varphi}{}_{s_n}\hspace{-2pt}}\hspace{2pt}$  is the component of $\sqrt{\hat{\varphi}{}_{s_n}\hspace{-2pt}}\hspace{2pt}$ associated to $\sqrt{\vphantom{\hat{\varphi}{}_{s_n}}\tilde{\varphi}{}_{s_n}\hspace{-2pt}}\hspace{2pt}$, i.e. the correctly estimated component,
${e}_{\varphi_{s_n}}^{\textsl{int\,}} = [\mathbf{e}_{\varphi_s} ^{\textsl{int\,}}]_n$
contains components associated to $\sqrt{\vphantom{\hat{\varphi}{}_{s_{n'}}}\tilde{\varphi}{}_{s_{n'}}\hspace{-2pt}}\hspace{2pt}$ with $n' \neq n$, i.e. erroneously estimated leakage or interference components across sources, and ${e}_{\varphi_{s_n}}^{\textsl{art\,}} = [\mathbf{e}_{\varphi_s} ^{\textsl{art\,}}]_n$ contains components not associated to any $\sqrt{\vphantom{\hat{\varphi}{}_{s_n}}\tilde{\varphi}{}_{s_{n}}\hspace{-2pt}}\hspace{2pt}$, i.e. erroneously estimated artifact components.
Exemplary spectrograms illustrating the decomposition in (\ref{eq:sim:decomp}) are shown in Fig. \ref{fig:spectrograms}, cf. also the discussion in Sec. \ref{sec:sim:acoustic:results}.

Given $L$ frames of $\sqrt{\vphantom{\hat{\bphi}_{s}}\bar{\bphi}_{s}}$ , $\mathbf{e}_{\varphi_s} ^{\textsl{int\,}}$ and $\mathbf{e}_{\varphi_s}^{\textsl{art\,}}$, we define the signal-to-interference ratio $\mathit{SIR}(\kappa)$, the signal-to-artifacts ratio $\mathit{SAR}(\kappa)$, and the signal-to-distortion ratio $\mathit{SDR}(\kappa)$ per third-octave band $\kappa$ along the lines of \cite{VincentGF06} as
\begin{align}
\mathit{SIR}(\kappa) &= 
10\log_{10}
\frac{\sum_{k,l}
\bigl\Vert \sqrt{\vphantom{\hat{\bphi}_{s}}\bar{\bphi}_{s}}(k,l)\bigr\Vert^2_2
}
{\sum_{k,l}
\bigl\Vert{\mathbf{e}_{\varphi_s}^{\textsl{int\,}}}(k,l)\bigr\Vert^2_2
}\,\si{dB},\\[2pt]
\mathit{SAR}(\kappa) &= 
10\log_{10}
\frac{\sum_{k,l}
 \bigl\Vert \sqrt{\vphantom{\hat{\bphi}_{s}}\bar{\bphi}_{s}}(k,l) + {\mathbf{e}_{\varphi_s}^{\textsl{int\,}}}(k,l)\bigr\Vert^2_2
}
{\sum_{k,l}
\bigl\Vert
{\mathbf{e}_{\varphi_s}^{\textsl{art\,}}}(k,l)\bigr\Vert^2_2
}\,\si{dB}, \\[2pt]
\mathit{SDR}(\kappa) &= 
10\log_{10}
\frac{\sum_{k,l}
\bigl\Vert \sqrt{\vphantom{\hat{\bphi}_{s}}\bar{\bphi}_{s}}(k,l)\bigr\Vert^2_2
}
{\sum_{k,l}
\bigl\Vert{\mathbf{e}_{\varphi_s}^{\textsl{int\,}}}(k,l) +
{\mathbf{e}_{\varphi_s}^{\textsl{art\,}}}(k,l)\bigr\Vert^2_2
}\,\si{dB},
\end{align}
with $k = k^{\scriptscriptstyle-}_\kappa,\dots,k^{\scriptscriptstyle+}_\kappa$ and $k^{\scriptscriptstyle-}_\kappa$ and $k^{\scriptscriptstyle+}_\kappa$  the frequency-bin indices of the lower and upper band limits of third-octave-band $\kappa$, and $l = 0,\dots,L-1$.

The decomposition in (\ref{eq:sim:decomp}) relies on a segment of $L$ frames of ${\tilde{\bphi}{}_{s}}$ and ${\hat{\bphi}{}_{s}}$ and is done in the following manner.
Let $\ushort{\hat{\bphi}}{}_{s_n}$ be a vector stacking the early PSD estimates ${\varphi}{}_{s_n}$ of source $n$ over $L$ observed frames, i.e.
$\ushort{\hat{\bphi}}{}_{s_n}  =
\begin{pmatrix}
\hat{\varphi}{}_{s_n}(0) &\cdots &\hat{\varphi}{}_{s_n}(L\-1)
\end{pmatrix}^\transp$,
and let $\ushort{\tilde{\bphi}}{}_{s_n}$, $\ushort{\bar{\bphi}}{}_{s_n}$, $\ushort{\mathbf{e}}_{\varphi_{s_n}}^{\textsl{int\,}}$ and $\ushort{\mathbf{e}}_{\varphi_{s_n}}^{\textsl{art\,}}$ be defined equivalently, such that
$\sqrt{\ushort{\tilde{\bphi}}{}_{s_n}\hspace{-2pt}}\hspace{2pt} = 
\sqrt{\vphantom{\ushort{\hat{\bphi}}{}_{s_n}}\ushort{\bar{\bphi}}{}_{s_n}\hspace{-2pt}}\hspace{2pt} +
\ushort{\mathbf{e}}_{\varphi_{s_n}}^{\textsl{int\,}} + 
\ushort{\mathbf{e}}_{\varphi_{s_n}}^{\textsl{art\,}}$,  similarly to (\ref{eq:sim:decomp}).
Then, we perform the orthonormal projection of each individual vector $\sqrt{\ushort{\hat{\bphi}}{}_{s_n}\hspace{-2pt}}\hspace{2pt}$ onto the one-dimensional subspace spanned by the corresponding vector $\sqrt{\vphantom{\ushort{\hat{\bphi}}{}_{s_n}}\ushort{\tilde{\bphi}}{}_{s_n}\hspace{-2pt}}\hspace{2pt}$, yielding $\sqrt{\vphantom{\ushort{\hat{\bphi}}{}_{s_n}}\ushort{\bar{\bphi}}{}_{s_n}\hspace{-2pt}}\hspace{2pt}$ 
with $\sqrt{\vphantom{\ushort{\hat{\bphi}}{}_{s_n}}\ushort{\bar{\bphi}}{}_{s_n}\hspace{-2pt}}\hspace{2pt} 	\propto \sqrt{\vphantom{\ushort{\hat{\bphi}}{}_{s_n}}\ushort{\tilde{\bphi}}{}_{s_n}\hspace{-2pt}}\hspace{2pt}$,
as well as onto the $N$-dimensional subspace spanned by all $N$ vectors $\sqrt{\ushort{\tilde{\bphi}}{}_{s_n}\hspace{-2pt}}\hspace{2pt}$, yielding $\sqrt{\vphantom{\ushort{\hat{\bphi}}{}_{s_n}}\ushort{\bar{\bphi}}{}_{s_n}\hspace{-2pt}}\hspace{2pt} + \ushort{\mathbf{e}}_{\varphi_{s_n}}^{\textsl{int\,}}$, which then allows us to explicitly compute  $\ushort{\mathbf{e}}_{\varphi_{s_n}}^{\textsl{int\,}}$ and $\ushort{\mathbf{e}}_{\varphi_{s_n}}^{\textsl{art\,}}$.
For further details, we refer the interested reader to \cite{VincentGF06}. 

\subsubsection{Acoustic Scenario}
\label{sec:sim:acoustic:acoustscen}

We use RIRs of $0.61\,$\si{s} reverberation time to a physical linear microphone array of $M = 5$ microphones with an inter-microphone distance of $8\,$\si{cm}  \cite{madivae}, similar to the assumed microphone array in Sec. \ref{sec:sim:modeled:datagen}.
We simulate $N = 2$ sources, using female and male speech \cite{bando92} as source signals.
The sources are assigned to two out of three possible source positions in $2\,$\si{m} distance of the microphone array at $\{-30,  0,  60\}^\circ$ relative to the broad-side direction, yielding six different speaker-source-position combinations. 
From the two source signal files, we randomly select $32$ segments of $5\,$\si{s} each.
Per segment-pair, we generate microphone signals for each speaker-source-position combination.

\subsubsection{Algorithmic Settings}
\label{sec:sim:acoustic:algorithmic}

In the acoustic-data case, the sampling frequency is $f_s = 16\,$\si{kHz}, and the STFT-analysis and synthesis is based on square-root Hann windows of $N_{\textsl{STFT}} = 512$ samples (corresponding to $32\,$\si{ms}) with $50\%$ overlap, resulting in $L = 312$ frames per segment. 
The desmoothed correlation matrix estimate $\hat{\bPsi}_x$ (cf. Sec. \ref{sec:smoothGEVD} and Sec. \ref{sec:desmoothGEVD}) is computed using $\zeta = e^{{-N_{\textsl{STFT}}}/{2f_s \tau}}$ with 
$\tau = 160\,$\si{ms}.
As in Sec. \ref{sec:sim:modeled:datagen}, $\bGamma$ is computed assuming a spherical-isotropic sound field. 
Given $\hat{\bPsi}_x$ and $\bGamma$,  we compute the estimates $\hat{\varphi}_\textsl{d}$, $\hat{\bPsi}{}_{x_\textsl{e}}$ and $\hat{\bPsi}{}^{\nicefrac{1}{2}}_{x_\textsl{e}}$ as described in Sec. \ref{sec:decompGEVD}.
We assume that the DoAs are known \cite{Thiergart2013AiMf, scheuing2008correlation, chen2010introduction}, and compute the (initial) estimate $\hat{\H}$ based on that.
Note that in a reverberant environment, where the free-field assumption does not hold, the RETFs are generally not only defined by the DoA, but also by early reflections, and therefore we generally have $\hat{\H} \neq \H$ in our setup.
Similarly to the model-based data case, cf. Sec. \ref{sec:sim:modeled:algo}, the penalty factor $\alpha$ in the conventional MP in (\ref{eq:sota:phixeHat_minprob}) and the square-root MP in (\ref{eq:oppPSD:minprob}) is simulated in the range $\alpha \in [10^{-3}, 10^{5}]$.
We perform at most $i_\textsl{max} = 20$ iterations of the associated iterative algorithms in (\ref{eq:sota:proxgradgrad})--(\ref{eq:sota:proxgradthres}) and (\ref{eq:oppPSD:minprobOmega})--(\ref{eq:oppPSD:minprobPhi}).
While we do not update  $\hat{\H}$ for the conventional MP in Sec. \ref{sec:sota}, we consider two cases for the square-root MP in Sec. \ref{sec:proposed}, namely first where we do not update $\hat{\H}$, and second where we update $\hat{\H}$ recursively as described in Sec. \ref{sec:RETFup}. 
In the latter case, in (\ref{eq:RETFup:xi}), since $\hat{\bPsi}{}^{\nicefrac{1}{2}}_{x_\textsl{e}}$  is subject to modeling and estimation errors and contains residual late reverberation, we set $\varphi_{\textsl{reg}} = \hat{\varphi}_{x_\ell}$.
In (\ref{eq:RETFup:w}), the threshold $\xi_\textsl{th}$ is again set as $10\log_{10} \xi_\textsl{th} = -2\,\si{dB}$ and $\beta$ is set per third-octave band $\kappa$ as $\beta(\kappa) = 20 \hat{b}^2(\kappa)$, with $\hat{b}(\kappa)$ pre-defined as the diversity of the Laplace distributions fitted to the real and imaginary parts of the STFT coefficients of a training signal within third-octave band $\kappa$.
Here, the training signal is generated from the entire female and male speech source signals, cf. Sec. \ref{sec:sim:acoustic:acoustscen}, by convolving the early part of the RIR of the first microphone corresponding to a source at $2\,\si{m}$ distance at $0^{\circ}$ relative to the broadside direction, cf. also the similar segment-wise definition of the reference signal $\tilde{s}_n$ in Sec. \ref{sec:sim:acoustic:measures}.
Note that while $\hat{b}(\kappa)$ is pre-computed using all STFT coefficients of both male and female speech within third-octave band $\kappa$, the actual distributions may vary across speakers, across source positions, across individual frequency bins, and across individual segments, cf. also Sec. \ref{sec:sim:acoustic:acoustscen}.

\subsubsection{Results}
\label{sec:sim:acoustic:results}

Before discussing the performance of the conventional MP and the square-root MP in terms of the measures $\mathit{SIR}$, $\mathit{SAR}$, and $\mathit{SDR}$, we first consider the examplary spectrograms in Fig. \ref{fig:spectrograms} visualizing the decomposition of $\sqrt{\hat{\bphi}{}_{s}}$ upon which these measures are based. 
In this example, the microphone signals $\x$ and the reference PSDs ${\tilde{\varphi}{}_{s_1}}$ and ${\tilde{\varphi}{}_{s_2}}$ 
originate from a female and a male speaker at $-30^\circ$ and $60^\circ$, respectively, and the estimates $\hat{\varphi}{}_{s_1}$ and $\hat{\varphi}{}_{s_2}$ in Fig. \ref{fig:spectrograms} (a.1) and Fig. \ref{fig:spectrograms} (a.2) are obtained by means of the square-root MP.
The correctly estimated components $\bar{\varphi}{}_{s_1}$ and $\bar{\varphi}{}_{s_2}$ in Fig. \ref{fig:spectrograms} (b.1) and Fig. \ref{fig:spectrograms} (b.2) are frequency-bin-wise scaled versions of the reference PSDs ${\tilde{\varphi}{}_{s_1}}$ and ${\tilde{\varphi}{}_{s_2}}$, respectively, cf. Sec. \ref{sec:sim:acoustic:measures}.
As can be seen, the leakage or interference components in ${e}_{\varphi_{s_1}}^{2|\textsl{int\,}}$ and ${e}_{\varphi_{s_2}}^{2|\textsl{int\,}}$ in  Fig. \ref{fig:spectrograms} (c.1) and  Fig. \ref{fig:spectrograms} (c.2) relate to the opposing reference PSDs, cf.  Fig. \ref{fig:spectrograms} (b.2) and Fig. \ref{fig:spectrograms} (b.1), respectively.
Finally, the artifact components ${e}_{\varphi_{s_1}}^{2|\textsl{art\,}}$ and ${e}_{\varphi_{s_2}}^{2|\textsl{art\,}}$ in  Fig. \ref{fig:spectrograms} (d.1) and  Fig. \ref{fig:spectrograms} (d.2) do not relate to any of the reference PSDs, but rather to residual late reverberation in the estimate $\hat{\bPsi}{}^{\nicefrac{1}{2}}_{x_\textsl{e}}$, cf. also Sec. \ref{sec:decompGEVD}, which is due to modeling errors in (\ref{eq:sm:Psiy})--(\ref{eq:sm:Psixl}) and a potential deviation of the late reverberant sound field from the spatial coherence matrix $\bGamma$.
Note that in ${e}_{\varphi_{s_1}}^{2|\textsl{art\,}}$ and ${e}_{\varphi_{s_2}}^{2|\textsl{art\,}}$, the energy is concentrated in the same spectro-temporal regions, indicating a similar spatial sound field of these components. 

Fig. \ref{fig:SIRSARSDR} shows the median over all segments and speaker-source-combinations, cf. Sec. \ref{sec:sim:acoustic:acoustscen}, of (a) $\mathit{SIR}$, (b) $\mathit{SAR}$, and (c) $\mathit{SDR}$ in third-octave bands for the conventional MP [\refConv], the square-root MP without recursive RETF update [\refFactNoUp], and the square-root MP with recursive RETF update [\refFactUp].
Here, in each third-octave band $\kappa$, we have selected $\alpha(\kappa)$ such that $\mathit{SIR}(\kappa)$ is maximized for each algorithm, i.e. the figure indicates their upper performance limit in terms of $\mathit{SIR}(\kappa)$ with respect to the tuning of $\alpha(\kappa)$.
Note that in our setup, selecting $\alpha(\kappa)$ to maximize $\mathit{SAR}(\kappa)$ or $\mathit{SDR}(\kappa)$ does not lead to qualitatively substantial differences.
For the conventional MP, we have found values of $\alpha(\kappa) \ll 1$ to be preferable in all third-octave bands $\kappa$, indicating that the soft-constraint penalty in (\ref{eq:sota:phixeHat_minprob}) is not very useful in practice. 
For the square-root MP, with and without recursive RETF update, we have found $\alpha(\kappa) \gg 1$ to be preferable in third-octave bands below $0.5\si{kHz}$, and $\alpha(\kappa) \leq 1$ to be preferable above $0.5\si{kHz}$.
From Fig. \ref{fig:SIRSARSDR} (a), we find that the square-root MP clearly outperforms the conventional MP in terms of $\mathit{SIR}$ in third-octave bands above $0.25\,\si{kHz}$, with improvements of $1\,\si{dB}$ to $6\,\si{dB}$, indicating better source-component separation performance.
Further, for the square-root MP, we find that the recursive RETF update mildly improves the performance by up to $1\,\si{dB}$.
Recall that the initial RETF estimate $\hat{\H}$ is based on the correct DoAs, but does not consider early reflections, cf. Sec. \ref{sec:sim:acoustic:algorithmic}.
From Fig. \ref{fig:SIRSARSDR} (b), we note that for all algorithms, we have $\mathit{SAR}(\kappa) < \mathit{SIR}(\kappa)$ in third-octave bands above $0.5\,\si{kHz}$, indicating comparably strong residual late reverberation.
The square-root MP performs slightly worse than the conventional MP in terms of $\mathit{SAR}$ in third-octave bands above $0.25\,\si{kHz}$, with degradations of less than $1\,\si{dB}$.
In the square-root MP, recursive RETF updating results in minor differences only.
As can be seen from Fig. \ref{fig:SIRSARSDR} (c), we find that the  square-root MP outperforms the conventional MP in terms of $\mathit{SDR}$, however, due to the comparably strong residual late reverberation, by much less than in terms of $\mathit{SIR}$.
Again, in the square-root MP, recursive RETF updating results in minor differences only.

 \setlength\fwidth{7.85cm}
 \begin{figure}[t]
\centering
\hspace*{-0.23cm}
    \setlength\fheight{8.25cm} 
    \input{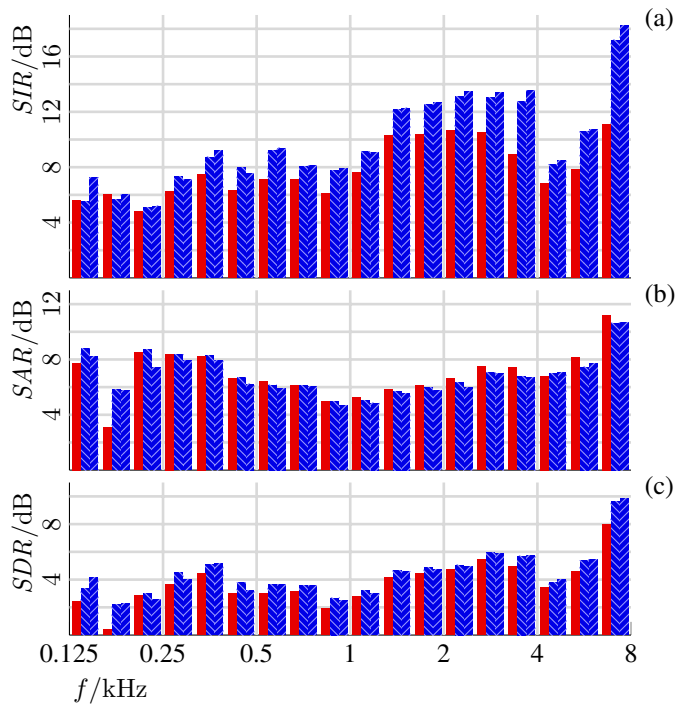} 
\caption{(a) $\mathit{SIR}$, (b) $\mathit{SAR}$, and (c) $\mathit{SDR}$ in third-octave bands for conventional MP [\refConv], square-root MP without recursive RETF update [\refFactNoUp], and square-root MP with recursive RETF update [\refFactUp].}
\label{fig:SIRSARSDR}
\end{figure}

\section{Conclusion}
\label{sec:conclusion}

We have discussed early PSD estimation in  the STFT domain for multiple sources in reverberant environments, based on a commonly used multi-microphone correlation matrix model, given (initial) RETF estimates.
State-of-the-art approaches to early PSD estimation formulate a minimization problem on the approximation error with respect to an estimate of the early correlation matrix, referred to as conventional MP.
Instead, we here have factorized the early correlation matrix model and formulated a corresponding minimization problem on the approximation error with respect to an estimate of the early-correlation-matrix square root, which we referred to as the square-root MP.
The square-root MP seeks a unitary matrix and the square roots of the early PSDs up to an arbitrary complex argument, and therewith constitutes a generalization of the orthogonal Procrustes problem.
As opposed to the conventional MP, non-negative inequality constraints are not required in the square-root MP.
The square-root MP may be solved iteratively, requiring one SVD per iteration.
Based on the estimated unitary matrix and early PSD square roots, we are further able to recursively update the RETF estimate, which is not inherently possible in the conventional approach.
The respectively required estimates of the early correlation matrix and the early-correlation-matrix square root may be obtained from an estimate of the microphone signal correlation matrix by means of the GEVD.
Hereat, in order to compensate for inevitable recursive averaging, we have restored non-stationarities by desmoothing the generalized eigenvlaues.

In order to evaluate the proposed approach, we have performed two kinds of simulations. 
In the first kind, the data is generated based on the microphone signal correlation matrix model and assumed geometric and physical properties, excluding modeling errors from the evaluation.
This is referred to as model-based-data case. 
In the second kind, the data is generated from recorded speech and measured RIRs, creating a more practical setup.
This is referred to as acoustic-data case.
In the model-based-data case, the simulation results indicate better performance of the square-root MP as compared to the conventional MP in terms of the relative squared PSD estimation error. 
If initialized accordingly, the square-root MP can be solved in only one iteration.
In the acoustic-data case, the simulation results indicate better performance of the square-root MP as compared to the conventional MP in terms of the source-component separation measured by the signal-to-interference ratio.
Both the square-root MP and the conventional MP suffer somewhat from residual late reverberation in the early-correlation-matrix estimate. 

\section*{Acknowledgments}
\label{sec:acknowledgements}

\footnotesize{This research work was carried out at the ESAT Laboratory of KU Leuven, in the frame of KU Leuven internal funds C2-16-00449, Impulse Fund IMP/14/037; IWT O\&O Project nr. 150611; VLAIO O\&O Project no. HBC.2017.0358; EU FP7-PEOPLE Marie Curie Initial Training Network funded by the European Commission under Grant Agreement no. 316969; the European Union's Horizon 2020 research and innovation program/ERC Consolidator Grant no. 773268. This paper reflects only the authors' views and the Union is not liable for any use that may be made of the contained information.}

\bibliographystyle{IEEEtran}
\bibliography{IEEEabrv}

\begin{thebibliography}{10}

\bibitem{beutelmann06}
R.~Beutelmann and T.~Brand,
\newblock ``Prediction of speech intelligibility in spatial noise and
  reverberation for normal-hearing and hearing-impaired listeners,''
\newblock {\em J. Acoust. Soc. Amer.}, vol. 120, no. 1, pp. 331–--342, Apr.
  2006.

\bibitem{loizou2007speech}
P.~C. Loizou,
\newblock {\em Speech enhancement: theory and practice},
\newblock CRC press, 2007.

\bibitem{benesty2008microphone}
J.~Benesty, J.~Chen, and Y.~Huang,
\newblock {\em Microphone array signal processing},
\newblock Springer, 2008.

\bibitem{gannot2017consolidated}
S.~Gannot, E.~Vincent, S.~Markovich-Golan, and A.~Ozerov,
\newblock ``A consolidated perspective on multimicrophone speech enhancement
  and source separation,''
\newblock {\em IEEE/ACM Trans. Audio, Speech, Lang. Process.}, vol. 25, no. 4,
  pp. 692--730, Apr. 2017.

\bibitem{quang04}
H.~Q. Dam, S.~Y. Low, H.~H. Dam, and S.~Nordholm,
\newblock ``Space constrained beamforming with source {PSD} updates,''
\newblock in {\em Proc. 2004 IEEE Int. Conf. Acoust., Speech, Signal Process.
  (ICASSP 2004)}, Montreal, QC, Canada, May 2004, pp. 93--96.

\bibitem{Thiergart2013AiMf}
O.~Thiergart, M.~Taseska, and E.~A.~P. Habets,
\newblock ``An informed parametric spatial filter based on instantaneous
  direction-of-arrival estimates,''
\newblock {\em IEEE/ACM Trans. Audio, Speech, Lang. Process.}, vol. 22, no. 12,
  pp. 2182--2196, Dec 2014.

\bibitem{BraunH15}
S.~Braun and E.~A.~P. Habets,
\newblock ``A multichannel diffuse power estimator for dereverberation in the
  presence of multiple sources,''
\newblock {\em {EURASIP} J. Audio Speech Music Process.}, vol. 2015, pp. 34,
  Dec. 2015.

\bibitem{SchwartzGH16}
O.~Schwartz, S.~Gannot, and E.~A.~P. Habets,
\newblock ``Joint maximum likelihood estimation of late reverberant and speech
  power spectral density in noisy environments,''
\newblock in {\em Proc. 2016 IEEE Int. Conf. Acoust., Speech, Signal Process.
  (ICASSP 2016)}, Shanghai, China, Mar. 2016, pp. 151--155.

\bibitem{Huang16}
Y.~A. {Huang}, A.~{Luebs}, J.~{Skoglund}, and W.~B. {Kleijn},
\newblock ``Globally optimized least-squares post-filtering for microphone
  array speech enhancement,''
\newblock in {\em Proc. 2016 IEEE Int. Conf. Acoust., Speech, Signal Process.
  (ICASSP 2016)}, Mar. 2016, pp. 380--384.

\bibitem{KuklasinskiDJJ16}
A.~Kuklasi\'nski, S.~Doclo, S.~H. Jensen, and J.~Jensen,
\newblock ``Maximum likelihood {PSD} estimation for speech enhancement in
  reverberation and noise,''
\newblock {\em {IEEE/ACM} Trans. Audio, Speech, Lang. Process.}, vol. 24, no.
  9, pp. 1599--1612, Sep. 2016.

\bibitem{SchwartzGH16b}
O.~Schwartz, S.~Gannot, and E.~A.~P. Habets,
\newblock ``Joint estimation of late reverberant and speech power spectral
  densities in noisy environments using {F}robenius norm,''
\newblock in {\em Proc. 24th European Signal Process. Conf. (EUSIPCO 2016)},
  Budapest, Hungary, Aug. 2016, pp. 1123--1127.

\bibitem{KodrasiIna2018JLRa}
I.~Kodrasi and S.~Doclo,
\newblock ``Joint late reverberation and noise power spectral density
  estimation in a spatially homogeneous noise field,''
\newblock in {\em Proc. 2018 IEEE Int. Conf. Acoust., Speech, Signal Process.
  (ICASSP 2018)}, Calgary, AB, Canada, Apr. 2018, pp. 441--445.

\bibitem{Braun18TASLP}
S.~{Braun}, A.~{Kuklasi\'nski}, O.~{Schwartz}, O.~{Thiergart}, E.~A.~P.
  {Habets}, S.~{Gannot}, S.~{Doclo}, and J.~{Jensen},
\newblock ``Evaluation and comparison of late reverberation power spectral
  density estimators,''
\newblock {\em IEEE/ACM Trans. Audio, Speech, Lang. Process}, vol. 26, no. 6,
  pp. 1056--1071, June 2018.

\bibitem{KodrasiD18}
I.~Kodrasi and S.~Doclo,
\newblock ``Analysis of eigenvalue decomposition-based late reverberation power
  spectral density estimation,''
\newblock {\em {IEEE/ACM} Trans. Audio, Speech, Lang. Process.}, vol. 26, no.
  6, pp. 1102--1114, June 2018.

\bibitem{Koutrouvelis18}
A.~I. Koutrouvelis, R.~C. Hendriks, R.~Heusdens, and J.~Jensen,
\newblock ``Robust joint estimation of multi-microphone signal model
  parameters,''
\newblock {\em IEEE/ACM Trans. Audio, Speech, Lang. Process.}, vol. 27, no. 7,
  pp. 1136--1150, July 2019.

\bibitem{jacobsen2000coherence}
F.~Jacobsen and T.~Roisin,
\newblock ``The coherence of reverberant sound fields,''
\newblock {\em J. Acoust. Soc. Amer.}, vol. 108, no. 1, pp. 204--210, July
  2000.

\bibitem{ephraim84}
Y.~Ephraim and D.~Malah,
\newblock ``Speech enhancement using a minimum mean-square error short-time
  spectral amplitude estimator,''
\newblock {\em IEEE Trans. Acoust., Speech, Signal Process.}, vol. 32, no. 6,
  pp. 1109--1121, Dec. 1984.

\bibitem{markovichAug09}
S.~Markovich, S.~Gannot, and I.~Cohen,
\newblock ``Multichannel eigenspace beamforming in a reverberant noisy
  environment with multiple interfering speech signals,''
\newblock {\em IEEE Trans. Audio, Speech, Lang. Process.}, vol. 17, no. 6, pp.
  1071--1086, Aug. 2009.

\bibitem{scheuing2008correlation}
J.~Scheuing and B.~Yang,
\newblock ``Correlation-based tdoa-estimation for multiple sources in
  reverberant environments,''
\newblock in {\em Speech and Audio Processing in Adverse Environments}, pp.
  381--416. Springer, 2008.

\bibitem{chen2010introduction}
Z.~Chen, G.~Gokeda, and Y.~Yu,
\newblock {\em Introduction to Direction-of-arrival Estimation},
\newblock Artech House, 2010.

\bibitem{everson98}
R.~Everson,
\newblock ``Orthogonal, but not orthonormal, {P}rocrustes problems,'' Tech.
  Rep., Laboratory for Applied Mathematics, City University New York and Mount
  Sinai Medical School, NYC, USA, 1998.

\bibitem{schonemann1966generalized}
P.~H. Sch{\"o}nemann,
\newblock ``A generalized solution of the orthogonal {P}rocrustes problem,''
\newblock {\em Psychometrika}, vol. 31, no. 1, pp. 1--10, Mar. 1966.

\bibitem{MantonJH2002}
J.~H. Manton,
\newblock ``Optimization algorithms exploiting unitary constraints,''
\newblock {\em IEEE Trans. Signal Process.}, vol. 50, no. 3, pp. 635--650, Mar.
  2002.

\bibitem{VincentGF06}
E.~Vincent, R.~Gribonval, and C.~F{\'{e}}votte,
\newblock ``Performance measurement in blind audio source separation,''
\newblock {\em {IEEE} Trans. Audio, Speech, Lang. Process.}, vol. 14, no. 4,
  pp. 1462--1469, 2006.

\bibitem{taslp19aCode}
T.~Dietzen,
\newblock ``Git{H}ub repository: square root-based multi-source early {PSD}
  estimation and recursive {RETF} update in reverberant environments by means
  of the orthogonal {P}rocrustes problem,''
  \url{https://github.com/tdietzen/SQRT-PSD-RETF}, July 2019.

\bibitem{avergel07}
Y.~Avargel and I.~Cohen,
\newblock ``System identification in the short-time {F}ourier transform domain
  with crossband filtering,''
\newblock {\em IEEE Trans. Audio, Speech, Lang. Process.}, vol. 15, no. 4, pp.
  1305--1319, Apr. 2007.

\bibitem{parikh2014proximal}
N.~Parikh and S.~Boyd,
\newblock ``Proximal algorithms,''
\newblock {\em Foundations and Trends in Optimization}, vol. 1, no. 3, pp.
  127--239, Jan. 2014.

\bibitem{antonello2018proximal}
N.~Antonello, L.~Stella, P.~Patrinos, and T.~van Waterschoot,
\newblock ``Proximal gradient algorithms: Applications in signal processing,''
  ESAT-STADIUS Tech. Rep. TR 17-112, KU Leuven, Belgium, Jan. 2018.

\bibitem{vanWaterschootToon2008Oraf}
T.~van Waterschoot, G.~Rombouts, and M.~Moonen,
\newblock ``Optimally regularized adaptive filtering algorithms for room
  acoustic signal enhancement,''
\newblock {\em Signal Processing}, vol. 88, no. 3, pp. 594--611, Mar. 2008.

\bibitem{LjungLennart1986Tapo}
L.~Ljung and T.~S{\"o}derstr{\"o}m,
\newblock {\em Theory and practice of recursive identification},
\newblock MIT press, 1986.

\bibitem{Serizel2014}
R.~Serizel, M.~Moonen, B.~Van~Dijk, and J.~Wouters,
\newblock ``Low-rank approximation based multichannel {W}iener filter
  algorithms for noise reduction with application in cochlear implants,''
\newblock {\em IEEE/ACM Trans. Audio, Speech, Lang. Process.}, vol. 22, no. 4,
  pp. 785--799, Apr. 2014.

\bibitem{MarkovichGolanShmulik2015Paot}
S.~Markovich-Golan and S.~Gannot,
\newblock ``Performance analysis of the covariance subtraction method for
  relative transfer function estimation and comparison to the covariance
  whitening method,''
\newblock in {\em Proc. 2015 IEEE Int. Conf. Acoust., Speech, Signal Process.
  (ICASSP 2015)}, Brisbane, QLD, Australia, Apr. 2015, pp. 544--548.

\bibitem{haykin02}
S.~Haykin,
\newblock {\em Adaptive Filter Theory},
\newblock Prentice-Hall, 4th edition, 2002.

\bibitem{dietzen19TASLP_ISCLP}
T.~Dietzen, S.~Doclo, M.~Moonen, and T.~van Waterschoot,
\newblock ``Integrated sidelobe cancellation and linear prediction {K}alman
  filter for joint multi-microphone dereverberation, interfering speech
  cancellation, and noise reduction,'' ESAT-STADIUS Tech. Rep. TR 19-70, KU
  Leuven, Belgium, submitted for publication, June 2019.

\bibitem{golub2012matrix}
G.~H. Golub and C.~F. Van~Loan,
\newblock {\em Matrix computations}, vol.~3,
\newblock Johns Hopkins University Press, Baltimore, MD, USA, 2012.

\bibitem{TammenMarvin2018CRoE}
M.~Tammen, I.~Kodrasi, and S.~Doclo,
\newblock ``Complexity reduction of eigenvalue decomposition-based diffuse
  power spectral density estimators using the power method,''
\newblock in {\em Proc. 2018 IEEE Int. Conf. Acoust., Speech, Signal Process.
  (ICASSP 2018)}, Calgary, AB, Canada, Apr. 2018, pp. 451--455.

\bibitem{Martin05}
M.~Martin,
\newblock ``Speech enhancement based on minimum mean-square error estimation
  and supergaussian priors,''
\newblock {\em {IEEE} Trans. Speech Audio Process.}, vol. 13, no. 5-2, pp.
  845--856, Sep. 2005.

\bibitem{LotterV05}
T.~Lotter and P.~Vary,
\newblock ``Speech enhancement by {MAP} spectral amplitude estimation using a
  super-gaussian speech model,''
\newblock {\em {EURASIP} J. Adv. Signal Process.}, vol. 2005, no. 7, pp.
  1110--1126, May 2005.

\bibitem{madivae}
E.~Hadad, F.~Heese, P.~Vary, and S.~Gannot,
\newblock ``Multichannel audio database in various acoustic environments,''
\newblock in {\em Proc. 2014 Int. Workshop Acoustic Signal Enhancement (IWAENC
  2014)}, Antibes -- Juan les Pins, France, Sept. 2014, pp. 313--317.

\bibitem{bando92}
Bang and Olufsen,
\newblock ``Music for {A}rchimedes,'' {C}ompact {D}isc B\&O, 1992.

\end{thebibliography}

\end{document}